\documentclass[12pt]{JHEP}
\usepackage{amsmath,amssymb}
\usepackage{ae,aecompl,cancel,comment}

\usepackage{yfonts}
\usepackage{pdfsync}
\usepackage{graphicx}

\newcommand{\be}{\begin{equation}}
\newcommand{\ee}{\end{equation}}
\newcommand{\beqa}{\begin{eqnarray}}
\newcommand{\eeqa}{\end{eqnarray}}

\def\bemat{\left( \begin{array}}
\def\enmat{\end{array} \right)}

\relax

\preprint{MPP-2011-63, DIAS-STP-11-03}

\title{Towards Unquenched Holographic Magnetic Catalysis}

\author{Veselin Filev \footnotemark[1]
\\Max-Planck-Institut f\"{u}r Physik (Werner-Heisenberg-Institut)
\\ F\"{o}hringer Ring 6, 80805 M\"{u}nchen, Germany
\vspace{.1cm}
\\\hspace{3.1cm}\&
\\School of Theoretical Physics
\\Dublin Institute for Advanced Studies
\\10 Burlington Road, Dublin 4, Ireland}

\author{Dimitrios Zoakos \footnotemark[2]
\\Centro de F\'\i sica do Porto \& Departamento de F\'\i sica e Astronomia
\\Faculdade de Ci\^encias da Universidade do Porto
\\Rua do Campo Alegre 687, 4169--007 Porto, Portugal}

\footnotetext[1]{E-mail address: \email{filev@mppmu.mpg.de}}
\footnotetext[2]{E-mail address: \email{dimitrios.zoakos@fc.up.pt}}

\abstract{We propose a string dual to the $SU(N_c)$ ${\cal N}=4$ SYM coupled to $N_f$ massless fundamental
flavors in an external magnetic field. The flavors are introduced by homogeneously smeared $N_f$ 
D7--branes and the external magnetic field via a non-trivial Kalb-Ramond $B$--field. 
Our solution is perturbative in a parameter that counts the number of internal flavor loops. 
In the limit of vanishing $B$--field the background reduces to 
the supersymmetric one obtained in hep-th/0612118. We introduce an additional probe D7--brane and in the
supersymmetric limit of vanishing $B$--field perform a holographic renormalization of its ``on-shell" action. 
We consider also non-supersymmetric probes with fixed worldvolume gauge field corresponding to a
magnetic field coupled only to the fundamental fields of the probe brane. We study the influence of the 
backreacted flavors on the effect of dynamical mass generation. Qualitatively the physical picture remains 
unchanged. In the next step we consider the case when the magnetic field couples to both the backreacted and 
the probe fundamental degrees of freedom. At sufficiently strong magnetic field the meson spectrum signals an 
instability of the probe D7--brane, which we interpret as reflecting an instability of the supergravity background.}

\begin{document}
\newpage
\section{Introduction}

In recent years an exciting area of theoretical physics has been unveiled through the discovery of the AdS/
CFT correspondence and its numerous generalizations, \cite{Maldacena:1997re}. 
Holographic techniques have proven a powerful analytic tool in studying the qualitative properties of 
strongly interacting physical systems with applications ranging from describing the physics of strongly 
interacting quark gluon plasmas to modeling collective condensed matter phenomena such as 
superconductivity and superfluidity.

Despite the success of the AdS/CFT correspondence in studying holographic gauge theories, namely field 
theories with known dual supergravity backgrounds, there are many realistic theories of great 
phenomenological importance, such as QCD, which do not have an explicit holographic dual.
This is the main reason of the limited direct quantitative applications of the correspondence. However, 
it turns out that under extreme external control parameters, such as: 
temperature, chemical potential and external 
electromagnetic fields, different gauge theories exhibit similar properties. Therefore it is natural to apply 
holographic techniques to study phenomena which are known to be of universal nature.

One such phenomenon is the magnetic catalysis of mass generation. At weak coupling it has been
extensively studied using the conventional perturbative field theoretic techniques 
\cite{perturbative}. 
A holographic study of this effect has been performed in \cite{Filev:2007gb}, 
where the case of flavored ${\cal N}=4$ Yang-Mills theory has been explored. 
Further holographic studies of this phenomenon have been addressed by numerous authors
\footnote{For a concise review in the case of holographic duals to the Dp/Dq--brane 
intersections look at ref.~\cite{Filev:2010pm}.} 
(for a comprehensive review of the literature look at ref.~\cite{Erdmenger:2007cm}). 
At present all such studies are in the limit of the so called ``quenched"  approximation when the number of 
fundamental fields $N_f$ is much smaller than the number of color degrees of freedom $N_c$.
On the gravity side this corresponds to the probe limit for the flavor D7-branes.

In this paper we undertake first steps towards a holographic description of the phenomenon of
magnetic catalysis of mass generation beyond the ``quenched" approximation. This suggests the 
construction of a supergravity background with a fully backreacted set of flavor branes coupled to a non-vanishing $B$-field. 
This construction amounts to a non-supersymmetric background.

Ideally such a background would correspond to a set of fully localized flavor branes.
The gravity dual of a field theory with unquenched flavor is coming through the solution of the equations of 
motion with brane sources. It is the presence of these sources which typically 
modify the Bianchi identities and through their contribution to the energy-momentum 
tensor, the Einstein equations also. If the flavor branes are 
localized, the sources contain Dirac delta functions and, as a consequence, solving the equations 
of motion is, in general, a difficult task. In the context of the AdS/CFT correspondence, the search for
localized solutions was initiated
in \cite{localized-I}, where the D3-D7 intersections were discussed,
while a lot of progress was reported in subsequent years, (see e.g.
\cite{localized-II}).

We could overcome the technical difficulties of localized flavor brane embeddings by considering 
configurations which are partially smeared
\footnote{For a detailed review on the smearing approach see \cite{Nunez:2010sf}.} 
along appropriately chosen compact directions. 
Supersymmetric backgrounds corresponding to partially smeared Karch-Katz like embeddings, 
\cite{Karch:2002sh},  have been constructed in \cite{Bigazzi:2008zt}.  In general these backgrounds 
possess a hollow cavity in the bulk of the geometry where the supergravity solution is sourced solely by the 
color branes. The radius of this cavity is related to the bare mass of the fundamental flavors \cite{Kruczenski:2003be}. In the limit of vanishing bare mass the cavity shrinks to the radius of the compact 
part of the geometry and the supergravity background has an essential singularity at the origin of the non-compact part of the geometry. In both cases the dilaton field diverges at large radial distances. This corresponds to the Landau pole that the dual field theory develops in the UV, due to its positive beta function~$\beta(\lambda)\propto N_f/N_c$.

One can imagine that a non-supersymmetric background interpolating between the supersymmetric 
backgrounds corresponding to massless flavors in the UV and massive flavors in IR would describe a 
dynamical mass generation. The radius of the hollow cavity would correspond to the dynamically generated 
constituent mass of the fundamental flavors.

A promising framework for the construction of such a geometry has been developed in 
\cite{Bigazzi:2009bk}.  In this paper, a ten dimensional black-hole solution dual to the 
non conformal plasma of ${\cal N}=4$ Yang-Mills coupled to $N_f\gg 1$ massless flavors 
has been presented
\footnote{All the hydrodynamic transport coefficients of the model were analyzed in \cite{hydro}, while the 
addition of a finite baryon density was presented in \cite{Bigazzi:2011it}} .
The D7 flavor branes are smeared homogeneously 
and extend along the radial direction up to the black hole horizon. 
The smearing procedure reduces the flavor symmetry group from $U(N_f)$ to $U (1)^{N_f}$ and allows 
for a simple way to account for the backreaction of the flavor branes. This in turn  allows one to explore the ``unquenched" regime of the dual field theory. In the zero temperature limit, the resulting backgrounds coincide with those found in~\cite{Benini:2006hh}.\footnote{Other solutions employing the smearing technique appear in~\cite{allsusyunquenched}.} More precisely, the authors of~\cite{Bigazzi:2009bk} consider the smearing of a general non-supesymmetric ``fiducial" embedding of the flavor brane for the purpose of obtaining a perturbative non-extremal black hole solution. 
However it turns out that obtaining even a perturbative solution in the general 
case is  technically too difficult and the authors obtain a perturbative finite temperature solution for the 
particular case of massless flavors (when the ``fiducial" embedding is trivial).

In this paper we will construct a perturbative non-supersymmetric background with a non-vanishing $B$-
field, which corresponds to an external magnetic field coupled to the fundamental degrees of freedom of the dual gauge theory. Following the approach of \cite{Bigazzi:2009bk} we will consider the case when the ``fiducial" embedding is trivial. Note that the results of the probe limit case considered in 
\cite{Filev:2007gb}, show that this embedding is unstable since it corresponds to vanishing constituent mass of the fundamental flavors. In the probe limit this instability is present for any non-vanishing value of the external magnetic field. Therefore one may expect that the background obtained by smearing such unstable embeddings would also be unstable. However the effect of backreaction is that the theory develops a Landau pole. This suggests the existence of an extra energy scale in addition to the energy scale associated to the external magnetic field. Therefore we may expect that the supergravity background would be characterized by a non-zero critical value of the $B$-field ($H_{cr}$) below which the background is stable. 
Our studies on the stability of the background using a probe D7-brane confirm this expectation.

Let us summarize the content of our work:

In Section 2 we introduce useful notation and the supergravity ansatz that is appropriate for the description of the
backreacted  supergravity background discussed above. We proceed by constructing an effective one 
dimensional action and obtaining the supergravity equations of motion. After that we solve those equations 
perturbatively to first order an appropriate small parameter, namely 
$\epsilon_*\propto \lambda_* \, \frac{N_f}{N_c}$. We are able to 
present our solution in a closed form. In the limit of vanishing magnetic field our solution reduces to the 
supersymmetric background constructed in \cite{Benini:2006hh}. 

In section 3 of this paper we introduce a probe flavor brane to the supersymmetric background obtained in\cite{Benini:2006hh}. We first consider the case of supersymmetric embeddings and perform a holographic regularization of the probe brane action. Our findings enable us to propose an AdS/CFT dictionary applicable at the finite UV cut off of the theory.  Next we consider D7-brane embeddings with fixed worldvolume gauge field. On the field theory side this corresponds to a constant external magnetic field coupled only to the fundamental degrees of freedom introduced by the probe brane. Using the proposed AdS/CFT dictionary we obtain numerical plot of the fundamental condensate versus the bare mass parameter and explore the effect that the presence of backreacted massless flavors have on the effect of magnetic catalysis of mass generation. Our findings suggest that the effect is to make the magnetic catalysis less efficient.

In Section 4 we generalize the study of Section 3 for an external magnetic field coupling to both the probe and the backreacted fundamental flavors. To this end, we consider a probe D7-brane in the non-supersymmetric background constructed in Section 2. We identify the value of the non-trivial $B$-field at the finite UV cut off as the value of the external magnetic field in the corresponding field theory. Our study of the classical embeddings of the probe brane suggest that for sufficiently low values of the perturbative parameter $\epsilon_*<\epsilon_{cr}$ the stable phase of the theory exhibit dynamical mass generation. However for sufficiently large values of $\epsilon_*=\epsilon_{cr}$ the theory is unstable. This instability is manifest as a diverging slope of the plot of the parameter of the IR separation (related to the constituent mass of the probe fundamental flavors) versus the parameter $\epsilon_*$. We find further evidence for this instability by studying the meson spectrum of the theory and identifying appropriate tachyonic modes. We interpret the instability of the probe as reflecting an instability of the supergravity background.

In the Conclusion section of the paper we discuss briefly our present results and their possible extensions. 


\section{Constructing the background}

The main goal we want to address with the gravity background of the following subsection is the 
study of the phenomenon of mass generation in magnetic catalysis. The field theories we are 
interested in are realized on the intersection between a set of $N_c$ {\it color} D3-branes and a 
set of $N_f$, homogeneously smeared, {\it flavor} D7-branes. This is mainly the construction that 
was studied in \cite{Bigazzi:2009bk} and in order to accommodate the phenomenon of 
mass generation we will impose 
an additional coupling between the fundamental fields and an external magnetic field.

The {\it color} D3-branes are placed at the tip of a Calabi-Yau (CY) cone over a 
Sasaki-Einstein manifold $X_5$, 
where the latter can be expressed as a $U(1)$ fiber bundle over a four dimensional 
K\"ahler-Einstein base (KE). In the absence of flavor branes the {\it color} ones source a background 
whose near horizon limit is the $AdS_5\times X_5$ and the dual gauge theories 
are superconformal quivers. 
\footnote{
For $X_5=S^5$ the CY is the six dimensional Euclidean space and the dual field theory 
is ${\cal N}=4$ SYM  \\ 
For $X_5=T^{1,1}$ the CY is the singular conifold and the dual theory is the Klebanov-Witten 
quiver \cite{Klebanov:1998hh}}

The {\it flavor} D7-branes introduce fundamental matter in the dual field theory. They extend along the 
radial direction of the background, wrap a submanifold $X_3$ of $X_5$ and at the same time smear  
homogeneously over the transverse space \cite{Casero:2006pt, Bigazzi:2005md}.
The smeared distribution is taken in such a way that the isometries of the fibered K\"ahler-Einstein
space are kept unbroken and allows to write an ansatz where all the unknown functions just depend on a 
single radial coordinate. The D7-brane embedding  is described by a constant profile, implementing 
massless flavor fields in the dual gauge theory.
As a general feature of all the D3-D7 setups, the dilaton runs and blows up at a certain radial distance, 
corresponding to a UV Landau pole in the dual gauge theory \cite{Benini:2006hh}.

The presence of both kind of branes will deform the ten dimensional
space-time, a product of a four dimensional Minkowski space with a six dimensional CY cone, 
through a (self dual) $F_5$ and an $F_1$ RR fields.  The coupling of the {\it flavor} D7-branes 
with the external magnetic field will be realized by the simultaneous presence of a $B_2$ NS field 
and its electric dual $C_2$ RR field, along the gauge theory directions.

In the next subsection we will be precise about the mathematical expression of 
every one of the forms that we described above.


\subsection{Ansatz}

Our notation will follow closely \cite{ Bigazzi:2009bk, Bigazzi:2011it, Benini:2006hh}. 
The action for the of Type IIB supergravity coupled to $N_f$ D7-branes, in the Einstein frame, is
given by the following expression 
\begin{equation}
S=S_{IIB} + S_{fl} \, , \label{genact}
\end{equation}
where the terms of the  $S_{IIB}$ action are
\begin{eqnarray} \label{TypeIIB action}
S_{IIB}&=&\frac{1}{2\kappa_{10}^2}\int d^{10} x \sqrt{-g} \Bigg[ R
- {1 \over 2} \partial_M\Phi \partial^M\Phi
- {1 \over 2} e^{2\Phi}F_{(1)}^2 
- {1 \over 2} \frac{1}{3!} e^{\Phi}  F_{(3)}^2
- {1 \over 2} \frac{1}{5!} F_{(5)}^2  \nonumber\\ \nonumber\\
&&\qquad \qquad \qquad \qquad \qquad  \quad \qquad \qquad  \,
- {1 \over 2} \frac{1}{3!} e^{-\Phi}  H_{(3)}^2\Bigg] 
- \frac{1}{2\kappa_{10}^2}\, \int C_4\wedge H_3\wedge F_{3} \, ,
\end{eqnarray}
and the action for the D7-branes takes the usual DBI+WZ form
\begin{equation}
S_{fl} = -T_7 \sum_{N_f} \Bigg[ \int d^8x\, e^\Phi 
\sqrt{-\det (\hat{G}+e^{-\Phi/2} \cal{F})}\, 
- \,\int_{D7}\hat C_q \wedge \left(e^{-\cal F}\right)_{8-q}  \Bigg] \, ,
\label{actionflav}
\end{equation}
with ${\cal F} \equiv  B + 2 \pi \alpha' F$. In those expressions $B$ denotes a non-constant magnetic field,
$F$ the worldvolume gauge field and the hat refers to the pullback of the quantities, 
along the worldvolume directions of the D7-brane.  The gravitational constant and D7-brane tension, 
in terms of string parameters, are
\begin{equation}
\frac{1}{2\kappa_{10}^2} = \frac{T_7}{g_s} = \frac{1}{(2\pi)^7g_s^2 \alpha'^4} \, .
\end{equation}
The ansatz for the metric that we will adopt is
inspired by \cite{Bigazzi:2009bk} and has the following form
\begin{equation}
ds_{10}^2 = h^{-\frac{1}{2}}\left[-dt^2 + dx_1^2+b(dx_2^2+dx_3^2)\right] + h^\frac{1}{2}
\left[ b^2 S^8F^2 d\sigma^2 + S^2 ds_{CP^2}^2 + F^2 (d\tau + A_{CP^2})^2 \right] \, ,
\label{10dmetric}
\end{equation}
where the $CP^2$ metric is given by 
\begin{eqnarray}
ds_{CP^2}^2&=&\frac{1}{4} d\chi^2+ \frac{1}{4} \cos^2 \frac{\chi}{2} (d\theta^2 +
\sin^2 \theta d\varphi^2) + \frac{1}{4} \cos^2 \frac{\chi}{2} \sin^2 \frac{\chi}{2}(d\psi + \cos \theta d\varphi)^2
\quad \& \nonumber \\
A_{CP^2}&=& \frac12\cos^2 \frac{\chi}{2}(d\psi + \cos \theta d\varphi)\,\,.
\label{cp2metric}
\end{eqnarray}
The range of the angles is $0\leq (\chi, \theta) \leq \pi$,  $0\leq \varphi, \tau < 2\pi$, $0\leq \psi< 4 \pi$.
The ansatz for the NS and the RR field strengths will be the following
\begin{eqnarray} \label{NS+RR}
& B_{2}  =  H \, dx^2\wedge dx^3 \, , \quad
C_{2} = J \, dt \wedge dx^1 \,,  & 
\nonumber \\ \nonumber \\
& F_{5}  =  Q_c\,(1\,+\,*)\varepsilon(S^5)\, , \quad
F_{1} = Q_f\,(d\tau + A_{CP^2})\, , \quad
F_{3} = d C_2\,+\, B_{2} \wedge F_1&
\end{eqnarray}
where $\varepsilon(S_5)$ is the volume element of the internal space and $Q_{c}, Q_{f}$ are proportional to
the number of colors and flavors
\begin{equation}
N_c = \frac{Q_c\, Vol(X_5)}{(2\pi)^4g_s \,\alpha'^2} \quad \& \quad
N_f = \frac{4\,Q_f\,Vol(X_5)}{Vol(X_3) g_s} \, .
\end{equation}
In our case $X_5=S^5$ and the volume of the three sphere is $2\pi^2$.
The fact that the flavors are massless is encoded in the independence  of $F_{(1)}$ on $\sigma$,
(see \cite{ Bigazzi:2008zt, Benini:2006hh}). All the functions that appear in the 
ansatz, $h,b,S,F, \Phi, J \, \& \, H$,
depend on the radial variable $\sigma$. In the convention we follow, $S \, \& \, F$ have dimensions of length, 
$b,h,J \, \& \, H$ are dimensionless and $\sigma$ has dimension length${}^{-4}$. 
The ansatz for the $F_3$ RR field strength is determined by the ansatz for $F_1$ (see Appendix A for more 
details). Finally the function $b$ in the ansatz for the metric reflects the breaking the of $SO(1,3)$ 
Lorentz symmetry down to $SO(1,1)\times SO(2)$.  

The equations of motion and Bianchi identities following from \eqref{genact} will be 
provided in full detail in Appendix A.

\subsection{Effective action and the equations of motion} \label{effectiveEOM}

Since all the functions that participate in the ansatz for the solution, 
\eqref{10dmetric} \& \eqref{NS+RR}, depend only on $\sigma$, 
it is feasible to describe the system in terms of a one-dimensional effective action.
In order to achieve that we insert all the ingredients of the ansatz in \eqref{genact} and integrate out
all the variables except $\sigma$ arriving to  
\begin{equation}
S_{eff}=\frac{\pi ^3 V_{1,3}}{2\kappa_{10}^2}\int {\cal L}_{1d} \, d\sigma
\end{equation}
where $V_{1,3}$ is the (infinite) volume of the Minkowski space and ${\cal L}_{1d} $ is given by the 
following expression
\begin{eqnarray} \label{L-effective1}
{\cal L}_{1d} &=&
-\frac{1}{2} \left(\frac{h'}{h}\right)^2 + 12 \left(\frac{S'}{S}\right)^2 + 8 \, \frac{F' S'}{F S}
+ 24\, b^2\, F^2\,S^6 - 4\, b^2\, F^4\,S^4
\nonumber \\ \nonumber \\
&+&\frac{b'}{b}\,\left( \frac{h'}{h}+ 8 \,\frac{S'}{S}+ 2\, \frac{F'}{F} \right)+  \frac{1}{2}\, 
\left(\frac{b'}{b}\right)^2 - \frac{b^2 Q_c^2}{2 h^2} -\frac{1}{2}\,Q_f^2 \,b^2e^{2\Phi} S^8 \,
\left(1+\frac{e^{-\Phi}\,H^2\,h}{b^2}\right) 
\\ \nonumber \\
&-&4\,Q_f\,b^2\,e^{\Phi}\,F^2\,S^6 \sqrt{1+\frac{e^{-\Phi}\,H^2\,h}{b^2}} - \frac{1}{2}\,\Phi'^2
- \frac{1}{2}\,\frac{e^{-\Phi}\,H'^2\,h}{b^2} \left(1 - \frac{e^{2\Phi}\,J'^2\,b^2}{H'^{2}}\right)
- Q_c H J' \, . \nonumber 
\end{eqnarray}
Producing \eqref{L-effective1} we have not made use of the WZ term, since it does not depend 
on the metric or the dilaton.  Its effect has been taken into account through the expression for
$F_{(1)}$ (see \cite{Benini:2006hh, Bigazzi:2009bk}). The precise expression of the WZ term is needed 
in producing the equations of motion explicitly from \eqref{genact} and it is this point that the 
smearing procedure enters the field and imposes the following replacement
\begin{equation}
\sum_{N_f}\int_{{\cal M}_8} \dots \quad \longrightarrow  \quad 
\int_{{\cal M}_{10}}\Omega\wedge\dots \ ,\label{smearing}
\end{equation}
where $\Omega$ is a form orthogonal to the D7-branes and we call it 
{\it smearing form}. The mathematical formula for the {\it smearing form} is 
given through the Bianchi identity for $F_{(1)}$ and it is
\begin{equation}
d F_1 = -  g_s \Omega_2\, .
\end{equation}
We will provide more details on the smearing of the DBI part of the action in Appendix A, where we 
will present all the details about the equations of motion and Bianchi identities of \eqref{genact}.
Since the potential $J$ enters the effective action only through its derivative, 
it corresponds to a ``constant of motion". This new parameter is
related to the value of the magnetic field close to the boundary through the
equations of motion for $F_3$, coming from the 10d supergravity. We will fix this
constant of motion in the following way
\begin{equation} \label{defJ}
\frac{\partial {\cal L}_{1d}}{\partial J'} \, \equiv  \, - \, Q_c H_{*} \quad \Rightarrow \quad 
J' \,=\, \frac{e^{-\Phi}\,Q_c}{h} \left(H \,-\,H_{*}\right)\, .
\end{equation}
The next step is to use equation \eqref{defJ} to eliminate
$J'$, in favor of $H_{*}$, from \eqref{L-effective1} after performing the following 
Legendre transformation 
\begin{equation}  \label{L-effective2}
\tilde {\cal L}_{1d} = L_{1d}-\frac{\delta L_{1d}}{\delta J'} \, J'\Bigg|_{J' \equiv J'(H_{*})} \, .
\end{equation}
The Euler-Lagrange equations will be calculated from the new, transformed,
action \eqref{L-effective2}. Defining the following auxiliary (dimensionless) expressions 
\begin{equation}
\beta_1 \equiv \sqrt{1+\frac{e^{-\Phi}\,H^2\,h}{b^2}} \, , \quad
\beta_2 \equiv 1 + \frac{e^{2\Phi}\,J'^{2}\,b^2}{H'^2} \quad \& \quad
\beta_3 \equiv 1 + \frac{e^{-2\Phi}\,H'^{2}\,\beta_2}{Q_f^2\,H^2\,b^2\,S^8}
\end{equation}
we can write the equations of motion in the following compact way
\begin{eqnarray}
\partial_\sigma^2(\log b)&=&-\, \frac{4 Q_f\,H^2\, h S^6 F^2}{\beta_1}
\,-\,e^{\Phi}\,H^2\,Q_f^2\,h \,S^8\,\beta_3 \label{diff-b}
\\ \nonumber \\
\partial_\sigma^2(\log h)&=&-Q_c^2 \frac{b^2}{h^2}\,-\,
\frac{2 Q_f\,H^2\, h S^6 F^2}{\beta_1}\,-\,
\frac{1}{2}\,e^{\Phi}\,H^2\,Q_f^2\,h \,S^8\,\beta_3\,
+\,\left(1-\beta_2\right)\,\frac{e^{-\Phi}\,h\,H'^2}{b^2} \label{diff-h}
\\ \nonumber \\
\partial_\sigma^2(\log S)&=& -2\, b^2 F^4 S^4 + 6\, b^2 F^2 S^6 - 
\frac{Q_f \,e^\Phi b^2 F^2\,S^6}{\beta_1} +
\frac{1}{4}\,e^{\Phi}\,H^2\,Q_f^2\,h \,S^8\, \beta_3 \label{diff-S}
\\ \nonumber \\
\partial_\sigma^2(\log F)&=& 4\,b^2 F^4 S^4 - 
\frac{1}{4}\,\left(1+\beta_1^2 \right)\,Q_f^2\,e^{2\Phi}b^2 S^8\,+\,
\frac{Q_f\,H^2\, h S^6 F^2}{\beta_1}
\,+\,\frac{1}{4}\,\frac{e^{-\Phi}\,h\,H'^2\,\beta_2}{b^2} \label{diff-F}
\\ \nonumber \\
\partial_\sigma^2\Phi&=& \frac{1}{2}\,\left(1+\beta_1^2 \right)
\left[Q_f^2\,e^{2\Phi}\,b^2\,S^8 \,+\, \frac{4 Q_f\,b^2\, e^\Phi S^6 F^2}{\beta_1}\right]\,
-\,\frac{1}{2}\,\frac{e^{-\Phi}\,h\,H'^2\,\beta_2}{b^2} \label{diff-Phi}
\\ \nonumber \\
\partial_{\sigma} \left[\frac{e^{-\Phi}\,h\,H'}{b^2}\right]&=& e^{\Phi}\,Q_f^2\,H\,h\,S^8\,+\,Q_c\,J'
+\, \frac{4 Q_f\,H\, h S^6 F^2}{\beta_1} \, . \label{diff-H}
\end{eqnarray}
It is straightforward to check that the above set of equations, together with \eqref{defJ}, 
solve the full set of Einstein equations, provided the following ``zero-energy'' constraint is also satisfied
\begin{eqnarray} \label{constraint}
0&=&
-\frac{1}{2} \left(\frac{h'}{h}\right)^2 + 12 \left(\frac{S'}{S}\right)^2 + 8 \, \frac{F' S'}{F S}
- 24\, b^2\, F^2\,S^6 + 4\, b^2\, F^4\,S^4
\nonumber \\ \nonumber \\
&+&\frac{b'}{b}\,\left( \frac{h'}{h}+ 8 \,\frac{S'}{S}+ 2\, \frac{F'}{F} \right)+  \frac{1}{2}\, 
\left(\frac{b'}{b}\right)^2 + \frac{b^2 Q_c^2}{2 h^2} + \frac{1}{2}\,Q_f^2 \,b^2e^{2\Phi} S^8\,\beta_1^2 
\\ \nonumber \\
&+&4\,Q_f\,b^2\,e^{\Phi}\,F^2\,S^6\,\beta_1 - \frac{1}{2}\,\Phi'^2
- \frac{1}{2}\,\frac{e^{-\Phi}\,h\, H'^{2}\,\beta_2}{b^2}\,+\frac{1}{2}\,e^{\Phi}\,h\,J'^2\, . 
\nonumber 
\end{eqnarray}
This constraint can be thought of as the $\sigma\sigma$ component of the Einstein equations.
Differentiating \eqref{constraint} and using \eqref{defJ} \& \eqref{diff-b}--\eqref{diff-H} we get zero,
meaning that the system is not overdetermined. 


\subsection{Perturbative solution}

The system \eqref{defJ} \& \eqref{diff-b}--\eqref{diff-H} allows for a perturbative solution along the lines 
of \cite{Bigazzi:2011it}. There, the dimensionless parameter $\epsilon_{*}$, 
which is  connected to the position that the dilaton blows up, 
was introduced and subsequently used as an expansion parameter.  
In order for the solution to be valid on a large energy 
scale this parameter has to be a small number and in principle the smaller the value of the number 
the larger the range of the energy scale.  After defining $\lambda_{*}$ 
as the 't Hooft coupling at the energy scale $r_{*}$ the parameter 
$\epsilon_{*}$ is expressed in terms of the physical quantities as
\begin{equation}
\epsilon_{*}\,=\, Q_f \, e^{\Phi_{*}} \, = \, 
\frac{Vol(X_3)}{16\pi\,Vol(X_5)} \, \lambda_{*}  \frac{N_f}{N_c} \, ,
\end{equation}
and particularly for our case that $X_5\,=\,S^5$ we have
$\epsilon_{*} = \frac{1}{8\pi^2}\lambda_*  \frac{N_f}{N_c}$. 
The parameter $\epsilon_{*}$ can be thought of as a flavor-loop counting
parameter in the dual field theory.

Obtaining a perturbative solution, in terms of $\epsilon_{*}$ of the system 
\eqref{defJ} \& \eqref{diff-b}--\eqref{diff-H}, 
we need to impose the following two requirements to fix the 
constants of integration in the order by order expansion of the solution. 
The first one is that the geometries
should coincide with the extremal ones in the absence of the magnetic field and the second is 
that the functions $F,S \, \& \, \Phi$ should correspond to the expressions given in 
\cite{Bigazzi:2011it}, at the energy scale $r=r_*$. We will redefine the radial variable $\sigma$ in such a way 
that the warp factor keeps the standard $AdS$ form
\begin{equation}
h=\frac{R^4}{r^4} \qquad \& \qquad R^4\equiv \frac14 Q_c
\end{equation}
therefore to first order in $\epsilon_{*}$ we have the following expression
\begin{eqnarray} \label{sigma-to-r}
\sigma &=&\frac{1}{4r^4} + \epsilon_{*} \, \Bigg[ -{1 \over 72\, r^4}
\left[ \frac{\alpha_{r}}{\alpha_{r}^2-1}-\frac{r^4}{r_{*}^4}\, 
\frac{\alpha_{r_{*}}}{\alpha_{r_{*}}^2-1}\right] +
 {1 \over 96\, r^4}
\left[ \alpha_{r} \left(\alpha_{r}^2-1\right) -\frac{r^4}{r_{*}^4}\, 
\alpha_{r_{*}}\left(\alpha_{r_{*}}^2-1\right)\right]
\nonumber \\ \nonumber \\
&-&  {1 \over 192\, r^4}
\left[\left(\alpha_{r}^2-1\right)^2\, \log \left[\frac{\alpha_r+1}{\alpha_r-1}\right]\ 
-\frac{r^4}{r_{*}^4}\, 
\left(\alpha_{r_{*}}^2-1\right)^2
\log \left[\frac{\alpha_{r_{*}}+1}{\alpha_{r_{*}}-1}\right] \right] - 
\frac{17}{144\,r^4} \,\left(\alpha_{r}-\alpha_{r_{*}}\right)
\nonumber \\ \nonumber \\
&-&  {1 \over 16\, r^4}
\left[\left(\alpha_{r}^2-1\right)\,
\log \left[\frac{\alpha_{r}+1}{\alpha_{r}-1}\right]
- \left(\alpha_{r_{*}}^2-1\right)
\log \left[\frac{\alpha_{r_{*}}+1}{\alpha_{r_{*}}-1}\right]
\right] + 
\frac{\alpha_{r_{*}}}{144 \,r^4}\, \left(1-\frac{r^4}{r_{*}^4}\right)\Bigg]\ ,
\end{eqnarray}
where another set of auxiliary dimensionless functions are in order
\begin{equation}
\alpha_r \equiv \sqrt{1+\frac{e^{-\Phi_*}\,H_{*}^2\,Q_c}{4\,r^4}} \quad \& \quad 
\alpha_{r_{*}} \equiv \sqrt{1+\frac{e^{-\Phi_{*}}\,H_{*}^2\,Q_c}{4\,r_{*}^4}} \, .
\end{equation}
Having \eqref{sigma-to-r} to transform the solution of the system \eqref{defJ} \& \eqref{diff-b}--\eqref{diff-H}  
from the one holographic coordinate to the other we arrive to the following expressions for $\Phi$, 
$H$, $J'$ and $b$
\begin{eqnarray}
\Phi &=& \Phi_{*}\,-\, \frac{\epsilon_{*}}{2} \, \Bigg[
\alpha_r - \alpha_{r_{*}} - \frac{1}{2}\, \log\left[
\frac{(\alpha_r +1)(\alpha_{r_{*}}-1)}{(\alpha_r -1)(\alpha_{r_{*}}+1)}\right]\Bigg] 
\label{Phi-r}
\\ \nonumber \\
H &=& H_{*}\Bigg[1\,-\, \frac{\epsilon_{*}}{8} \, 
\left[\alpha_r \,\frac{\alpha_r^2+1}{\alpha_r^2-1}\, 
-\, \alpha_{r_{*}}\,\frac{\alpha_{r_{*}}^2+1}{\alpha_{r_{*}}^2-1}\, \frac{r^4}{r_{*}^4}\,
-\, \frac{\alpha_r^2-1}{2}\, \log \left[\frac{\alpha_r+1}{\alpha_r-1}\right]\, \right.
\label{H-r}
 \\ \nonumber \\
&& \qquad \qquad \qquad \qquad \qquad \qquad \qquad \qquad \qquad
\left.+\, \frac{\alpha_{r_{*}}^2-1}{2}\, \log \left[\frac{\alpha_{r_{*}}+1}{\alpha_{r_{*}}-1}\right]
\, \frac{r^4}{r_{*}^4}\, \right]\Bigg]
\nonumber \\ \nonumber \\
J' &=& \epsilon_{*}\,\frac{e^{-\Phi_{*}}\,H_{*}}{2\,r}\,\Bigg[
\alpha_r \,\frac{\alpha_r^2+1}{\alpha_r^2-1}\,
-\,\alpha_{r_{*}} \,\frac{\alpha_{r_{*}}^2+1}{\alpha_{r_{*}}^2-1}\, \frac{r^4}{r_{*}^4}\,
-\,\frac{\alpha_r^2-1}{2}\, \log \left[\frac{\alpha_r+1}{\alpha_r-1}\right]\,
\label{Jp-r}
 \\ \nonumber \\
&& \qquad \qquad \qquad \qquad \qquad \qquad \qquad \qquad \qquad \,\,\,
+\,\frac{\alpha_r^2-1}{2}\, \log \left[\frac{\alpha_{r_{*}}+1}{\alpha_{r_{*}}-1}\right]
\, \frac{r^8}{r_{*}^8}\, \Bigg]
 \nonumber \\ \nonumber \\
b &=& 1\,+\, \frac{\epsilon_{*}}{2} \, \Bigg[ \alpha_r- \alpha_{r_{*}} 
+\frac{1}{2}\,\left(\alpha_r^2-1\right)\log \left[\frac{\alpha_{r}+1}{\alpha_{r}-1}\right]
- \frac{1}{2}\,\left(\alpha_{r_{*}}^2-1\right)
\log \left[\frac{\alpha_{r_{*}}+1}{\alpha_{r_{*}}-1}\right]\Bigg] \, . \label{b-r}
\end{eqnarray}
Integrating \eqref{constraint} we have for $F$ \& $S$ 
\begin{eqnarray}
F+ 4 S &=& 5r + \frac{\epsilon_{*}}{2} \, \Bigg[ -\frac{r}{16}\,
\left[ \alpha_{r}\left(\alpha_{r}^2-1\right)-\frac{r^4}{r_{*}^4}\, \alpha_{r_{*}}
\left(\alpha_{r_{*}}^2-1\right)\right]+\frac{r}{9}\frac{r^4}{r_{*}^4} \nonumber \\
&+&\frac{r}{8}\,
\left[ \alpha_{r}-\frac{r^4}{r_{*}^4}\, \alpha_{r_{*}} \right]+
\frac{r}{4}\,
\left[ \frac{\alpha_{r}}{\alpha_{r}^2-1}-\frac{r^4}{r_{*}^4}\, 
\frac{\alpha_{r_{*}}}{\alpha_{r_{*}}^2-1}\right] \\
&+& \frac{r}{32}\left[ 
\left(\alpha_r^2-1\right)^2
\log \left[\frac{\alpha_{r}+1}{\alpha_{r}-1}\right]
-\frac{r^4}{r_{*}^4}\, 
 \left(\alpha_{r_{*}}^2-1\right)^2
 \log \left[\frac{\alpha_{r_{*}}+1}{\alpha_{r_{*}}-1}\right]
\right]\Bigg] \nonumber 
\end{eqnarray}
while decoupling \eqref{diff-S} \& \eqref{diff-F} 
\begin{eqnarray}
F - S &=& -\, \frac{\epsilon_{*}}{12} \, \Bigg[ \frac{r}{4}\,
\left[ \alpha_{r}\left(\alpha_{r}^2-1\right)-\frac{r^2}{r_{*}^2}\, \alpha_{r_{*}}
\left(\alpha_{r_{*}}^2-1\right)\right]+ r\,\frac{r^2}{r_{*}^2} 
-\frac{r}{4}\,\left(\alpha_{r}^2-1\right)^{3/2}\, \left(1-\frac{r^8}{r_{*}^8}\right)  \label{F-S}\nonumber \\
&+&\frac{5 \, r}{8}\,
\left[ \alpha_{r}-\frac{r^2}{r_{*}^2}\, \alpha_{r_{*}} \right]+
\frac{3}{8}\,\frac{r}{\sqrt{\alpha_{r}^2-1}}\,
\log\left[\frac{\alpha_r+\sqrt{\alpha_{r}^2-1}}{\alpha_{r_{*}}+\sqrt{\alpha_{r_{*}}^2-1}}\right]\Bigg] \, .\label{f-s}
\end{eqnarray}


\subsection{Validity of the perturbative solution}

The perturbative solution we present in the preceding section needs to be supplemented with a
hierarchy of scales. In terms of the radial coordinate $r$ (or $\sigma$) our solution has two 
pathological regions that we will describe in this section both qualitatively \& quantitatively.

As explained in the introduction, the background we have constructed consists of a sea of
massless flavors. This means that the flavor branes are stretched down to the bottom of the 
geometry and the charge density is highly peaked at $r=0$. This is the reason behind the 
curvature singularity that appears in the origin.\footnote{A very nice 
visualization of this construction can be found in \cite{Nunez:2010sf}.}
Another way to understand this pathology is through the restoration of the $U(N_f)$ symmetry 
in the deep IR and the subsequent 
increase of the effective string coupling in such a way that the smearing approach breaks down. 
Careful choice of the integration constants when obtaining the perturbative solution 
together with the presence of the 
magnetic field, produce a less severe IR singularity but by no means cure it.

A crucial point is that due to the infrared divergency of the background the perturbative solution presented 
in equations \eqref{sigma-to-r}--\eqref{F-S} is not valid below certain radial distance above 
the origin. In particular the Jacobean $\Big|\frac{\partial{ \sigma}}{\partial r}\Big|$ is 
vanishing at some $ r_{IR}(\epsilon_*)$. As one may expect if we increase the number of backreacted 
flavors the value for the radius $r_{IR}(\epsilon)$ increases. One can imagine that higher order 
corrections in $\epsilon_*$ could reduce the value $r_{IR}(\epsilon)$, however the intrinsic infrared 
divergency of the background at the origin suggests that $r_{IR}$ remains non-vanishing in any 
perturbative solution. In Sections 3 and 4 we will use this radius as a parameter characterizing the infrared 
applicability of our probe-brane analysis.

There are two ways of either avoiding or hiding the infrared singularity. The first one is to pull the flavor branes 
away from the origin, while keeping the radial symmetry, by introducing massive flavors. Then the distance 
between the color and the flavor branes is interpreted as a mass for the fundamentals. The second 
possibility is the addition of temperature to the background, which has effect of hiding the singularity behind 
the horizon, see \cite{Bigazzi:2009bk}.

A common characteristic of all the unquenched holographic backgrounds constructed so far is the 
appearance of a point in the deep UV that the dilaton diverges. This behavior is a signal for a Landau 
pole in the dual gauge theory.\footnote{The first counter example is the addition of flavor D6-branes to the 
ABJM \cite{Conde:2011}, 
where the smeared unquenched supergravity solution has a {\it good} UV behavior.} 
The perturbative solution we constructed in the previous section is valid until $r=r_{*}$, 
where $r_{*}$ denotes an (arbitrary) UV cutoff scale $\frac{r_{*}}{R^2}\sim \Lambda_{UV}$,
well below the energy scale of the Landau pole.
Additionally the energy scale of $r_{*}$ should be well above the IR scale, in order for the 
UV completion to have only negligible effects in the IR physics. 
The main characteristic of this scale is that the UV details of the theory do not affect the 
IR physical predictions. This feature is reflected by the independence of all physical quantities 
(up to suppressed contributions) on the position of $r_{*}$ but only on IR parameters.
The quantitative outcome
of the above analysis is that $\epsilon_{*} \ll 1$, \cite{Bigazzi:2009bk, Bigazzi:2011it},
and as we increase the number of
flavors the {\it effective physical region} for the solution between the IR \& UV energy scales 
decreases.


\section{Probing the supersymmetric background}

In this section we introduce a probe D7-brane to the supersymmetric background of \cite{Benini:2006hh}, 
which is the limit of vanishing magnetic field for the perturbative supergravity background constructed above. 
We consider both supersymmetric and non supersymmetric embeddings. The latter have a fixed value of the 
$U(1)$-gauge field strength along the coordinates $x_2$ and $x_3$, namely $F_{23}={\rm const}$. On the 
field theory side this corresponds to an external magnetic field coupling only to the fundamental degrees of 
freedom introduced by the probe D7-brane. We will contrast those results to the ones of Section 4, where the 
external magnetic field couples to all the fundamental degrees of freedom. 
Let us begin by studying the properties of the supersymmetric embeddings. 

 
\subsection{Holographic renormalization}

In this subsection we propose a holographic renormalization scheme similar to the one reported in
\cite{Karch:2005ms} for asymptotically AdS gravitational backgrounds. Our goal is to come up with a 
prescription to calculate vacuum expectation values (like the fundamental condensate) for the supergravity 
background of \cite{Benini:2006hh}, which suffers from UV divergencies. In the limit of vanishing magnetic 
field the first order perturbative solution is given by \cite{Bigazzi:2009bk}
\begin{eqnarray}
\Phi&=&\Phi_*+\epsilon_*\log\frac{r}{r_*}+{\cal O}\left(\epsilon_*^2\right) \, , \quad \quad b=0 
\, , \quad \quad h=\frac{R^4}{r^2} \\
F_0&=&r \Bigg[1-\epsilon_* \, \frac{1}{24}\left(1+\frac{1}{3}\frac{r^4}{r_*^4}\right)\Bigg]+
{\cal O}\left(\epsilon_*^2\right) \quad \& \quad 
S_0=r \Bigg[1+\epsilon_*\frac{1}{24}\left(1-\frac{1}{3}\frac{r^4}{r_*^4}\right)\Bigg]+
{\cal O}\left(\epsilon_*^2\right) \, .
\nonumber 
\end{eqnarray}
The lagrangian of the D7-brane probe is given by
\begin{equation}
-\frac{{\cal L}}{\cal N}=\frac{e^{\Phi} }{8r^5}S_0^6F_0^2\cos^3{\frac{\chi}{2}}\left(\cos^2\frac{\chi}{2}+\frac{S_0^2}{F_0^2}\sin^2\frac{\chi}{2}\right)^{\frac{1}{2}}\left(1+\frac{r^{10}\chi'^2}{4S_0^6F_0^2}\right)^{\frac{1}{2}}+\frac{\epsilon_*e^{2\Phi-\Phi_*}}{32r^5}S_0^8\cos^4\frac{\chi}{2} \, , 
\label{Lagr-SUSY}
\end{equation}
where ${\cal N} \equiv 8\,T_7 \,{\rm Vol}(S^3)\, V_{1,3}$ with $V_{1,3}$ the volume spanned by ($t,\,x^i$) and
${\rm Vol}(S^3)=2\pi^2$ the volume of the unit $S^3$ wrapped by the probe brane. In the next step we expand 
the classical embedding around $\epsilon_*$ according to $\chi=\chi_0(r)+\epsilon_*\chi_1(r)$ and solve perturbatively the EOM obtained from (\ref{Lagr-SUSY}). The zeroth order lagrangian is given by
\begin{equation}
-\frac{{\cal L}_0}{\cal N}= \frac{1}{8}e^{\Phi_*}r^3\cos^3\frac{\chi_0(r)}{2}\left(1+\frac{r^2}{4}\chi_0'(r)^2\right)^{1/2}\ ,
\end{equation}
which is just the lagrangian for a D7-brane probe in pure AdS$_5\times S^5$ space time analyzed in 
\cite{Karch:2002sh} (in slightly different coordinates). The solution for a supersymmetric embedding 
corresponding to flavor with bare mass $m/(2\pi\alpha')$ is given by
\begin{equation}
\chi_0(r)=2\arcsin\frac{m}{r}\ . \label{0themb}
\end{equation}
The solution for $\chi_1(r)$ corresponding to the first order correction to (\ref{0themb}) is
%
%
\begin{equation}
\chi_1(r)=\frac{m\left[r^4-m^4+12\,r_*^4 \log({m \over r})\right]}
{36\, r_*^4 \sqrt{r^2-m^2}} .\label{1themb}
\end{equation}
In the next step we substitute \eqref{0themb} and \eqref{1themb} in the lagrangian \eqref{Lagr-SUSY}, 
obtain the ``on-shell" lagrangian ${\cal L}_{cl}$ to first order in $\epsilon_*$
\begin{eqnarray}
-\frac{{\cal L}_{cl}}{\cal N} & = & \frac{1}{8}e^{\Phi_*}r(r^2-m^2) +
\frac{\epsilon_*  \, e^{\Phi_*}}{288 \, r \, r_*^4}
\Bigg[-12 \, m^2 \, r^2 \, r_*^4\, \log\left({m \over r}\right) 
\\ 
&-& \left(r^2-m^2\right) \left[m^4\,r^2 + 4\, r^6 - 15 \, r^2 \, r_*^4 + m^2\,  \left( r^4 + 3 r_*^4 \right)
- 36 r^2 r_*^4 \log\left({r \over r_*}\right) \right]\Bigg] \nonumber 
\end{eqnarray}
and then integrate it from $m$ to $r_*$. Note that since $\chi_1(m)=0$ the classical 
embedding of the D7-brane, to first order, closes at $r=m$ above the origin. It is convenient to define
\begin{equation}
{S}_{cl}=\int\limits_m^{r_*}dr{\cal L}_{cl}=S_{cl}^{(0)}+\epsilon_*S_{cl}^{(1)}+
{\cal O}\left(\epsilon_*^2\right) \, ,
\end{equation}
and after that expand in powers $r_*$
\begin{eqnarray}
-\frac{S_{cl}^{(0)}}{{\cal N}e^{\Phi_*}}&=&\frac{1}{32}(r_*^2-m^2)^2=\frac{r_*^4}{32}
-\frac{r_*^2m^2}{16}+\frac{m^4}{32}\ ,\label{classaction}
\\
-\frac{S_{cl}^{(1)}}{{\cal N}e^{\Phi_*}}&=&\frac{r_*^4}{288}
-\frac{r_*^2 \, m^2}{576} \, \left[5+12\log\left({m \over r_*}\right)\right]
+\frac{m^4}{192}\, \left[1+4\log\left({m \over r_*}\right)\right]+O\left(\frac{1}{r_*^2}\right)\ .
\end{eqnarray}
The divergent and finite terms can be canceled \cite{Karch:2005ms} by the addition of the following counter terms at $r=r_*$
\begin{eqnarray}
L_1&=&\#_1e^{\Phi}\sqrt{-\gamma} \, , \quad
L_2=\#_2e^{\Phi}\sqrt{-\gamma}\chi^2 \quad \& \quad 
L_f=\#_fe^{\Phi}\sqrt{-\gamma}\chi^4 \, ,\label{CTs}
\end{eqnarray}
where $\#_1,\#_2$ and $\#_f$ are appropriately chosen coefficients and $\gamma$ is the determinant of the 
induced metric on the $r={\rm const}$ slice. Note that we have integrated along all the compact directions 
(their contribution is included in ${\cal N}$) and  we have added a factor of $e^{\Phi}$ to all the counter terms. 
To first order in $\epsilon_*$ we have the following divergent and finite contributions from the counter terms
\begin{eqnarray}
\frac{L_1}{\#_1e^{\Phi_*}}&=&\frac{r_*^4}{R^4} \, ,\label{CT1}
\\
\frac{L_2}{\#_2e^{\Phi_*}}&=&\frac{4r_*^2m^2}{R^4}+\frac{4m^4}{3R^4}
+\epsilon_* \, \frac{m^2}{9 \, R^4}\,\left(r_*^2 + \frac{2}{3}\,m^2 \right) \, 
\left[1+12\,\log\left({m \over r_*}\right)\right] \, , \label{CT2}
\\
\frac{L_f}{\#_fe^{\Phi_*}}&=&\frac{16m^4}{R^4}
+\epsilon_* \, \frac{8\,m^4}{9 \, R^4}\, 
\left[1+12\,\log\left({m \over r_*}\right)\right] \, . \label{CT3}
\end{eqnarray}
One can check that the following set of coefficients
 \begin{eqnarray}
\#_1&=&{\cal N} \, \frac{R^4}{32}\left(1+\frac{1}{9}\epsilon_*\right) \label{coeff1} \, , \\ 
\#_2&=& -\, {\cal N} \, \frac{R^4}{64}\left(1+\frac{1}{9}\epsilon_*\right) \label{coeff2}\, ,  \\
\#_f&=& {\cal N} \, \frac{5 \, R^4}{1536}\left(1+\frac{1}{9}\epsilon_*\right) \label{coeff3}
\end{eqnarray}
cancels all the divergent and finite terms in the first order of the ``on-shell" action 
$S_{cl}^{(0)}+\epsilon_*S_{cl}^{(1)}$.

An interesting observation coming from \eqref{coeff1}, \eqref{coeff2} and \eqref{coeff3} is that the 
coefficient of all the counter terms has the same $\epsilon_*$ dependence, namely  
$\left(1+\epsilon_*/9\right)$. In fact one can check that this remains true also for the second order 
corrections, suggesting that resummation is possible. Indeed, it turns out that if one considers the 
non-perturbative form of the solution the $\epsilon_*$ dependence of the ``on-shell" action factorizes. 
Let us briefly sketch the derivation. The supersymmetric background is determined by 
the following set of first order differential equations \cite{Bigazzi:2009bk}
\begin{eqnarray}
S&=&\alpha'^{\frac{1}{2}}e^{\rho}\Bigg[1+\epsilon_*(\frac{1}{6}+\rho_*-\rho)\Bigg]^{\frac{1}{6}}\, ,
\label{SS-S}  \\
F&=&\alpha'^{\frac{1}{2}}e^{\rho}\left[1+\epsilon_*(\rho_*-\rho)\right]^{\frac{1}{2}}
\Bigg[1+\epsilon_*\left(\frac{1}{6}+\rho_*-\rho\right)\Bigg]^{-\frac{1}{3}}\ ,
\\
\Phi&=&\Phi_*-\log\left[1+\epsilon_*(\rho_*-\rho)\right] \, , 
\\
\frac{dh}{d\rho}&=&-Q_c\alpha'^{-2}e^{-4\rho}
\Bigg[1+\epsilon_*\left(\frac{1}{6}+\rho_*-\rho\right)\Bigg]^{-\frac{2}{3}}\ ,\label{SS-h}
\end{eqnarray}
where a new radial coordinate $\rho$ has been introduced. The relation between $r$ and the 
new radial variable $\rho$ can be obtained as an expansion in powers series of $\epsilon^*$ 
and to first order is given by \cite{Bigazzi:2009bk}
\begin{equation}
r=\alpha'^{\frac{1}{2}}e^{\rho}\Bigg[1+\frac{\epsilon_*}{72}\,\left[e^{4\rho-4\rho*}-1+12(\rho_*-\rho)\right]\Bigg]
+{\cal O} \left(\epsilon_*^2\right)\, .\label{r-to-rho}
\end{equation}
Note that for $\epsilon_*=0$ one has simply $r=\alpha'^{\frac{1}{2}}e^{\rho}$, while this relation also holds at 
$r=r_*$ for arbitrary~$\epsilon_*$. Therefore at $\epsilon_*=0$ the supersymmetric embedding 
$\chi(\rho)$ corresponding to $\chi_0(r)$ from \eqref{0themb} is
\begin{equation}
\chi(\rho)\equiv 2\arcsin\frac{e^{\rho_q}}{e^\rho}\ .\label{susyemb}
\end{equation}
The parameter $e^{\rho_q}$ is related to the bare mass of the fundamental degrees of freedom introduced by 
the probe. It turns out that the supersymmetric embedding at finite $\epsilon_*$ is still 
given by equation (\ref{susyemb})\footnote{Note that equation (\ref{r-to-rho}) suggests that $m\equiv r(\rho_q)=\alpha'^{\frac{1}{2}}e^{\rho_q}+O(\epsilon)$.}, see \cite{Bigazzi:2009bk}. The next step is to evaluate the ``on-shell" action
$S_{cl}$ and for this we note the zeroth order term in \eqref{classaction} which suggests the following
$-S_{cl}^{(0)}/({\cal N}\alpha'^2)=e^{\Phi_*}(e^{2\rho_*}-e^{2\rho_q})^2/32$. The classical action to all orders in $\epsilon_*$ has 
the following expression\footnote{We refer the reader to the appendix of the paper for more details on 
this calculation.}
\begin{equation}
-\frac{S_{cl}(\epsilon_*)}{{\cal N}\alpha'^2}= \frac{e^{\Phi_*}}{32}\left(1+\frac{\epsilon_*}{6}\right)^{\frac{2}{3}}
\left (e^{2\rho_*} -e^{2\rho_q} \right)^2 \, . \label{exactcl}
\end{equation}
Remarkably the $\epsilon_*$ dependence of the ``on-shell" action factorizes when it is written in the radial coordinate $\rho$. Furthermore, since at $r=r_*$ 
the relation between the two radial coordinates is $r_*=\alpha'^{\frac{1}{2}}e^{\rho_*}$
the $\epsilon_*$ dependence of the ``on-shell" action remains unchanged if one defines $m_0=\alpha'^{\frac{1}{2}}e^{\rho_q}=m+O(\epsilon_*)$ as a bare mass parameter. This suggests that to all orders in 
$\epsilon_*$ the counter terms are given by equation \eqref{CTs} with the following coefficients 
$\#_1,\#_2,\#_3$
\begin{eqnarray}
\#_1& = & {\cal N} \, \frac{R^4}{32}\left(1 + \frac{1}{6}\epsilon_*\right)^{\frac{2}{3}} \, , 
\\
\#_2 & = & -\, {\cal N} \, \frac{R^4}{64}\left(1 + \frac{1}{6}\epsilon_*\right)^{\frac{2}{3}} \, ,
\\
\#_f & = & {\cal N} \, \frac{5 \, R^4}{1536}\left(1 + \frac{1}{6}\epsilon_*\right)^{\frac{2}{3}} \,  . 
\end{eqnarray}
The above considerations suggest that one can consistently regularize the ``on-shell" action of the probe 
D7-brane provided one keeps the parameter $\epsilon_*$ sufficiently small to ensure that the finite cut off 
$r_*=\alpha'^{\frac{1}{2}}e^{\rho_*}$ is sufficiently far from the Landau pole of the theory and at the same time is sufficiently 
large so that terms of order ${\cal O}(e^{-\rho_*})$ can be ignored in calculating the regularized ``on-shell" 
action. Therefore we propose that the usual AdS/CFT dictionary can be applied at $r=r_*$, 
namely one can expand
\begin{equation}
\sin\frac{\chi(r_*)}{2}=\frac{m_0}{r_*}+\frac{c}{r_*^3}+O\left(\frac{1}{r_*^5}\right)\label{dictionary}
\end{equation}
and identify $m_0=\alpha'^{\frac{1}{2}}e^{\rho_q}$ as the bare mass parameter and $c\propto \langle\bar qq\rangle$  as the 
fundamental condensate. Note that for the supersymmetric embedding of \eqref{susyemb} 
we have $c=0$ and thus the fundamental condensate vanish (as it should for supersymmetric theory). In the next subsection we will consider non-supersymmetric embeddings with fixed $U(1)$ gauge field strength $F_{23}={\rm const}$ and apply the proposed dictionary to study the fundamental condensate of the theory. 


\subsection{D7-brane probe with fixed $U(1)$-gauge field.}

In this subsection we introduce a D7-brane probe with fixed $U(1)$ gauge field strength 
$F_{23}=H_0/2\pi\alpha'$. On the field theory side this corresponds to coupling the ${\cal N}=2$ 
hypermultiplet with a constant external magnetic field $H_0/2\pi\alpha'$ along the $x_1$ direction. 
The effective action of the probe brane in the Einstein frame is
\begin{equation}
{\cal S}=-T_7\int d^8xe^{\Phi}\sqrt{-det(\hat g+e^{-\Phi/2}B_2)}+T_7\int P[ C_{(8)}]\ ,
\end{equation}
where $C_{(8)}$ is the background Ramond-Ramond form sourced by the smeared flavor branes. 
One can show that the effective lagrangian is
\begin{eqnarray}
-\frac{{\cal L}}{\cal N}&=& \frac{1}{8}e^{\Phi}S^6F^2\cos^3\frac{\chi}{2}\left(\cos^2\frac{\chi}{2}+
\frac{S^2}{F^2}\sin^2\frac{\chi}{2}\right)^{1/2}\left(1+\frac{\chi'^2}{4S^6F^2}\right)^{1/2}
\left(1+e^{-\Phi}H_0^2h\right)^{1/2}
\nonumber \\
&+&\frac{Q_f}{32}e^{2\Phi}S^8\cos^4\frac{\chi}{2}\ , \label{effact}
\end{eqnarray}
where the functions $F\, S\,\Phi$ \& $h$ are given in equations \eqref{SS-S}-\eqref{SS-h}. Note that 
in \eqref{SS-h} the equation of motion for $h$ is in an integral form but it can be obtained in 
terms of incomplete gamma functions. We refer the reader to the appendix for the technical details. 

\begin{figure}[h] 
   \centering
   \includegraphics[width=3.2in]{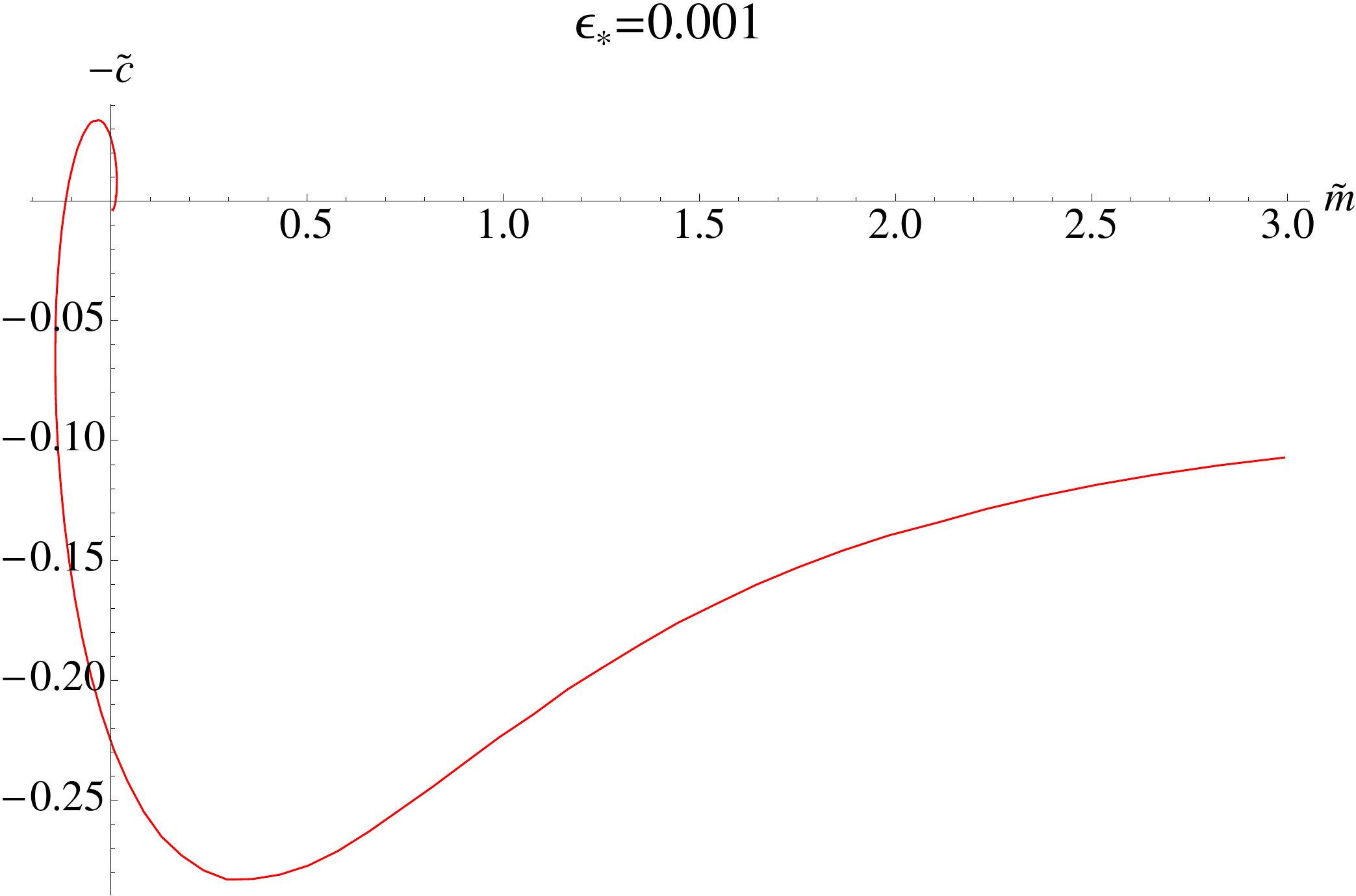} 
   \includegraphics[width=3.2in]{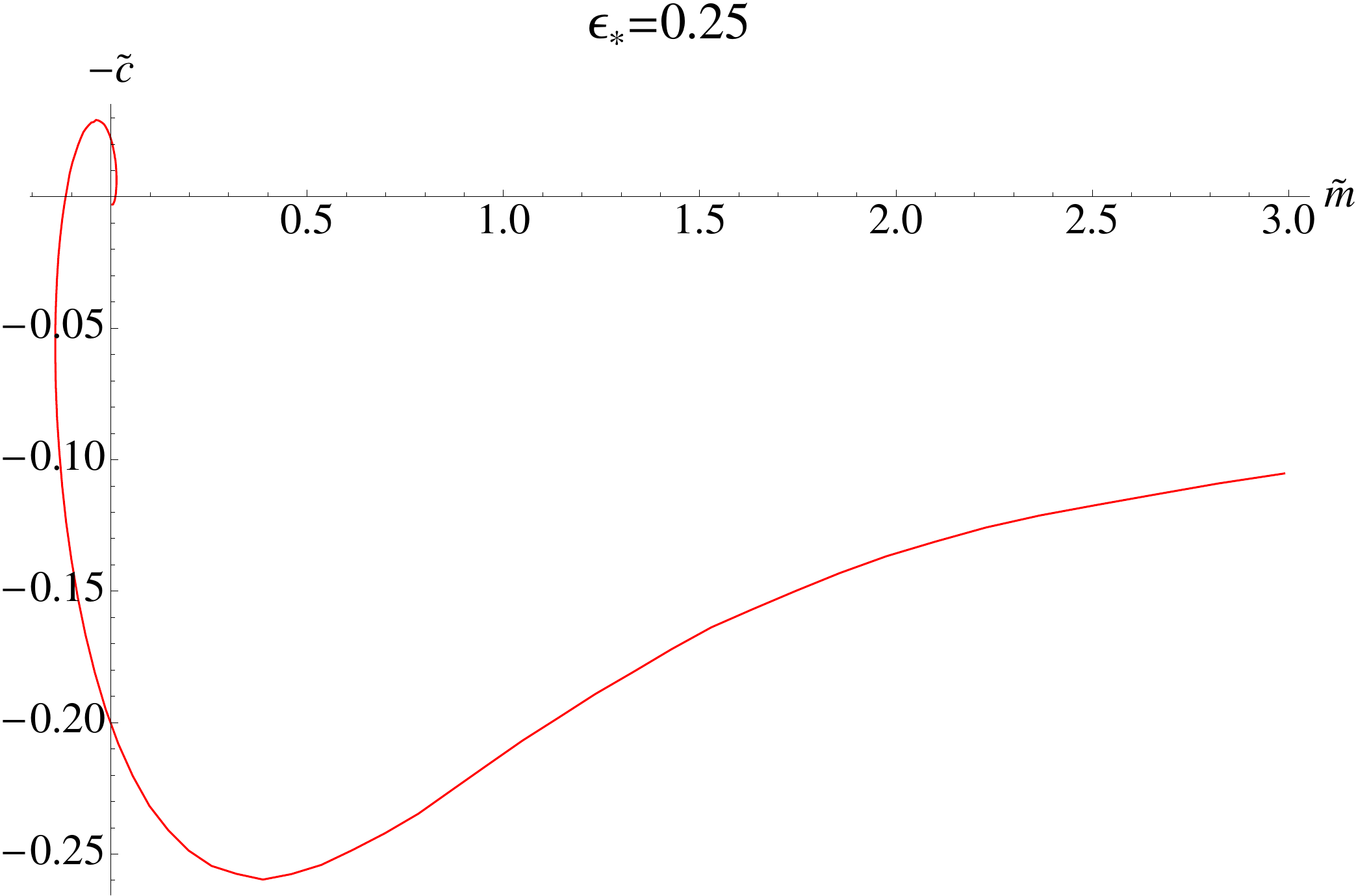} 
   \includegraphics[width=3.2in]{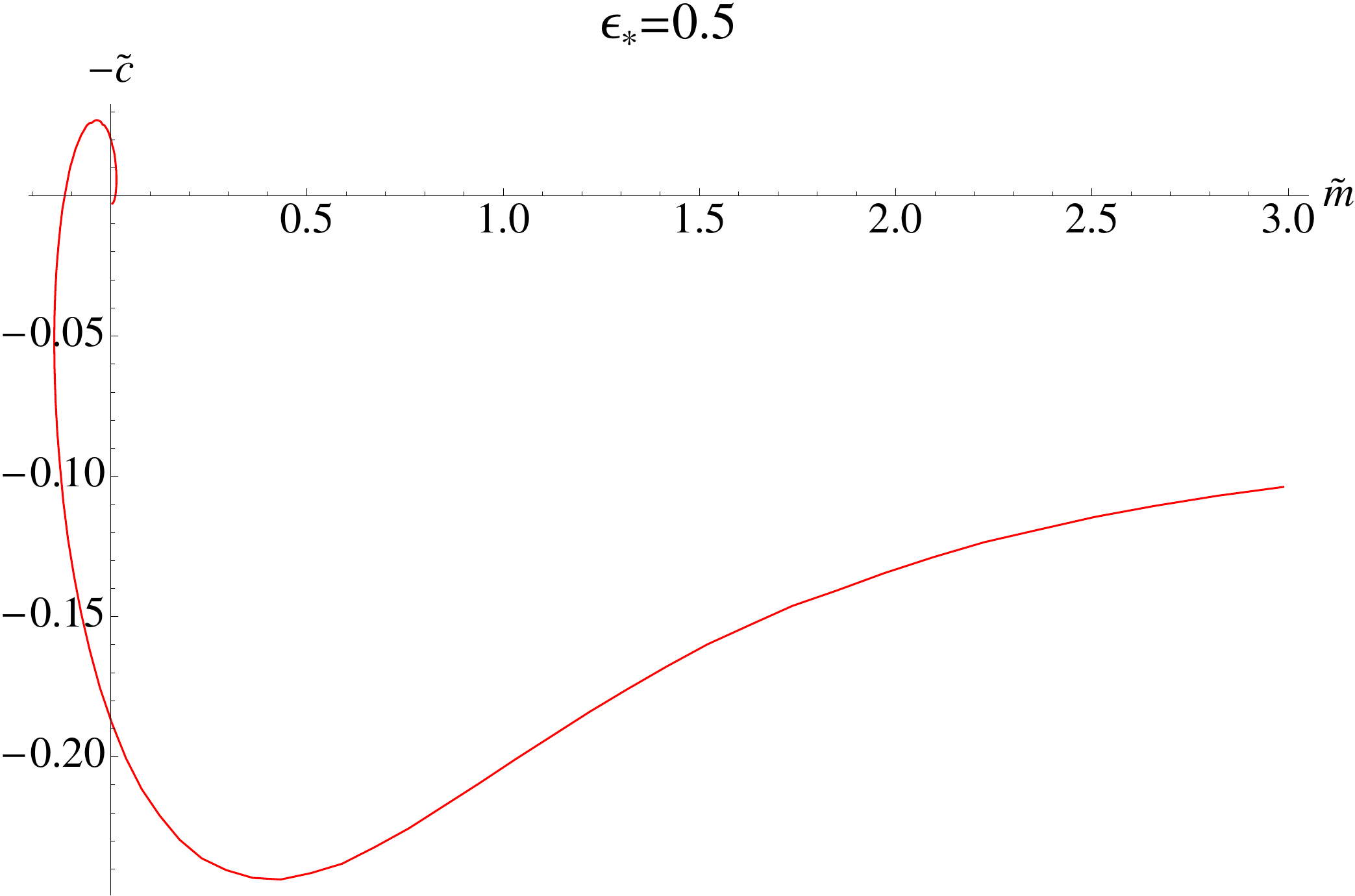} 
   \includegraphics[width=3.2in]{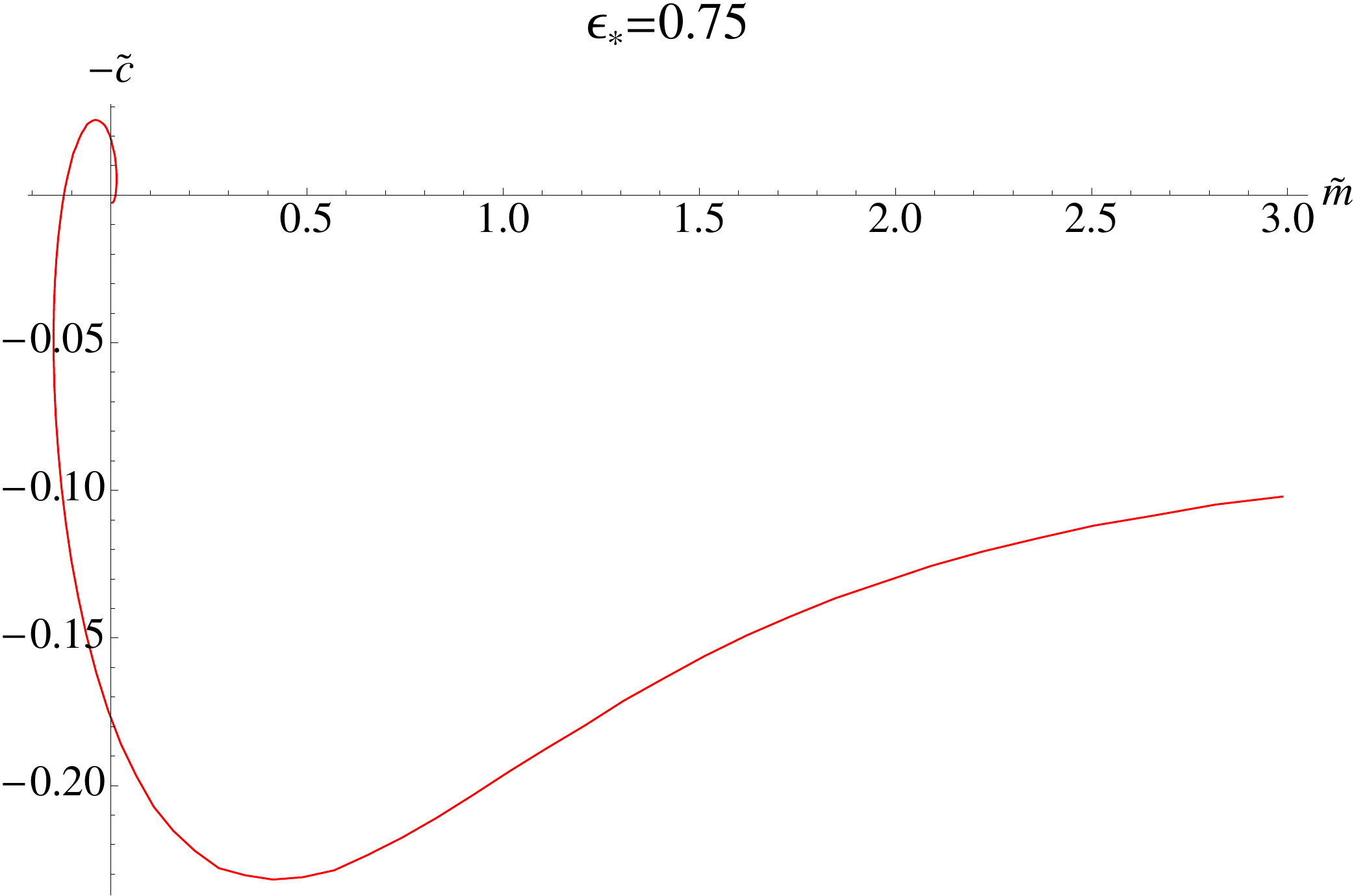} 
   \caption{\small Plots of the parameter $-\tilde c$ (fundamental condensate) versus the parameter $\tilde m$ (bare mass) for various values of the parameter $\epsilon_*$. The shape of the plots remains unchanged.}
   \label{fig:-1}
\end{figure}

It turns out that for the numerical analysis it is convenient to define the following dimensionless quantities
\begin{equation}
\tilde\rho=\rho-\rho_m \, , \quad 
\tilde r=r/r_m \, , \quad \& \quad 
\rho_m\equiv\log\frac{r_m}{\sqrt{\alpha'}}\equiv \frac{1}{4}\log\frac{e^{-\Phi_*}Q_c H_0^2}{4\alpha'^2}\, .
\label{changevar}
\end{equation}
Next we proceed by solving numerically the equation of motion for the classical embedding of the probe $\chi(\tilde\rho)$ and expanding at $\tilde\rho=\tilde\rho_*=\log(\tilde r_*)$
\begin{equation}
\sin\frac{\chi(\tilde r_* )}{2}=\frac{\tilde m}{\tilde r_*}+\frac{\tilde c}{r_*^3}+
{\cal O} \left(\frac{1}{\tilde r_*^5}\right) \quad {\rm with} \quad \tilde 
m=m_0/r_m \quad \& \quad \tilde c=c/r_m^3;\ , 
\end{equation}
where as we discussed above $m_0$ \& $c$ are related to the bare mass and the fundamental condensate of 
the ${\cal N}=2$ hypermultiplet corresponding to the probe brane.
Next we vary the parameter $\epsilon_*$ and generate plots of $-\tilde c$ versus $\tilde m$, which we 
present in figure \ref{fig:-1}. 

\begin{figure}[h] 
   \centering
   \includegraphics[width=3.2in]{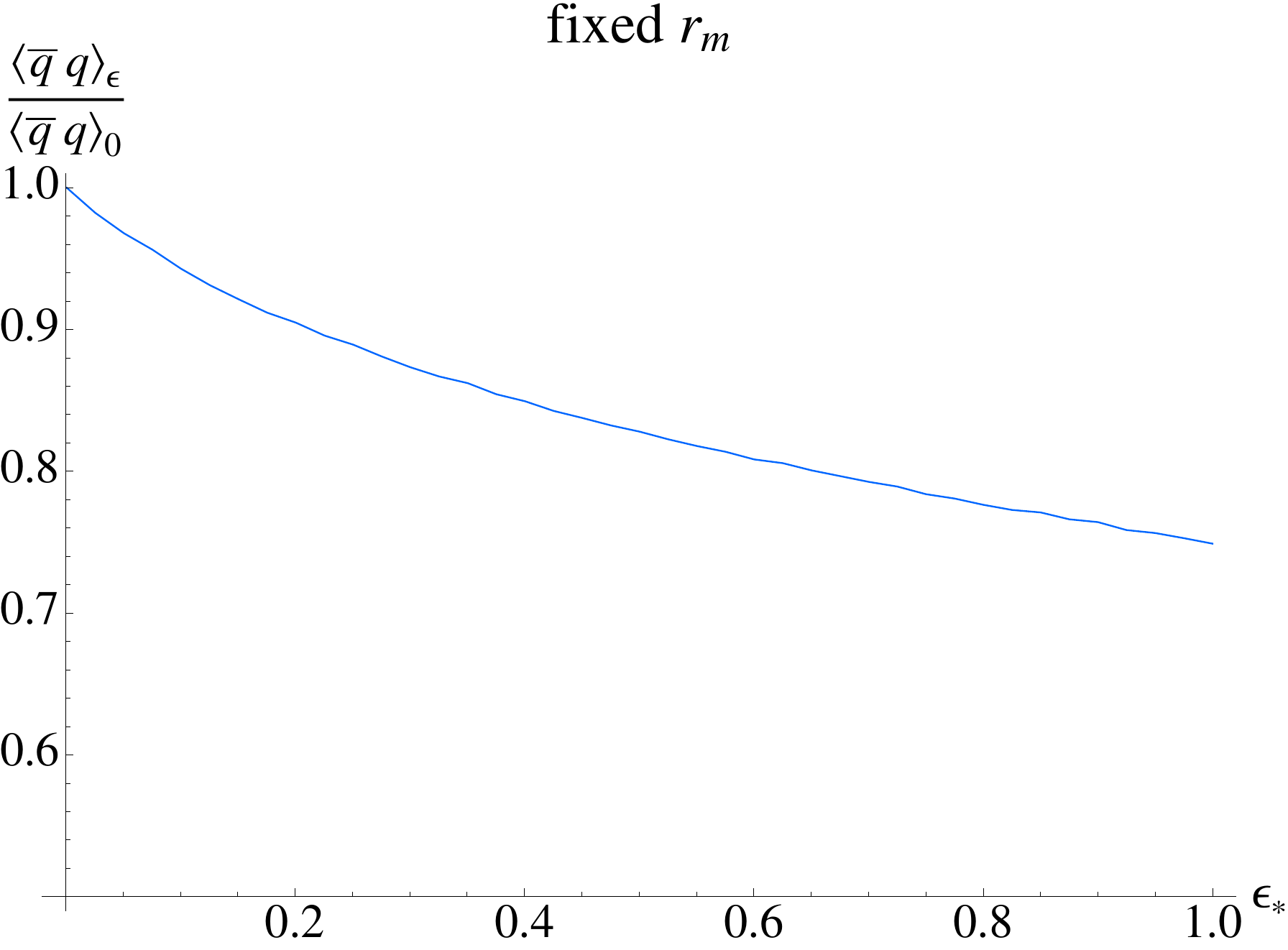} 
   \includegraphics[width=3.2in]{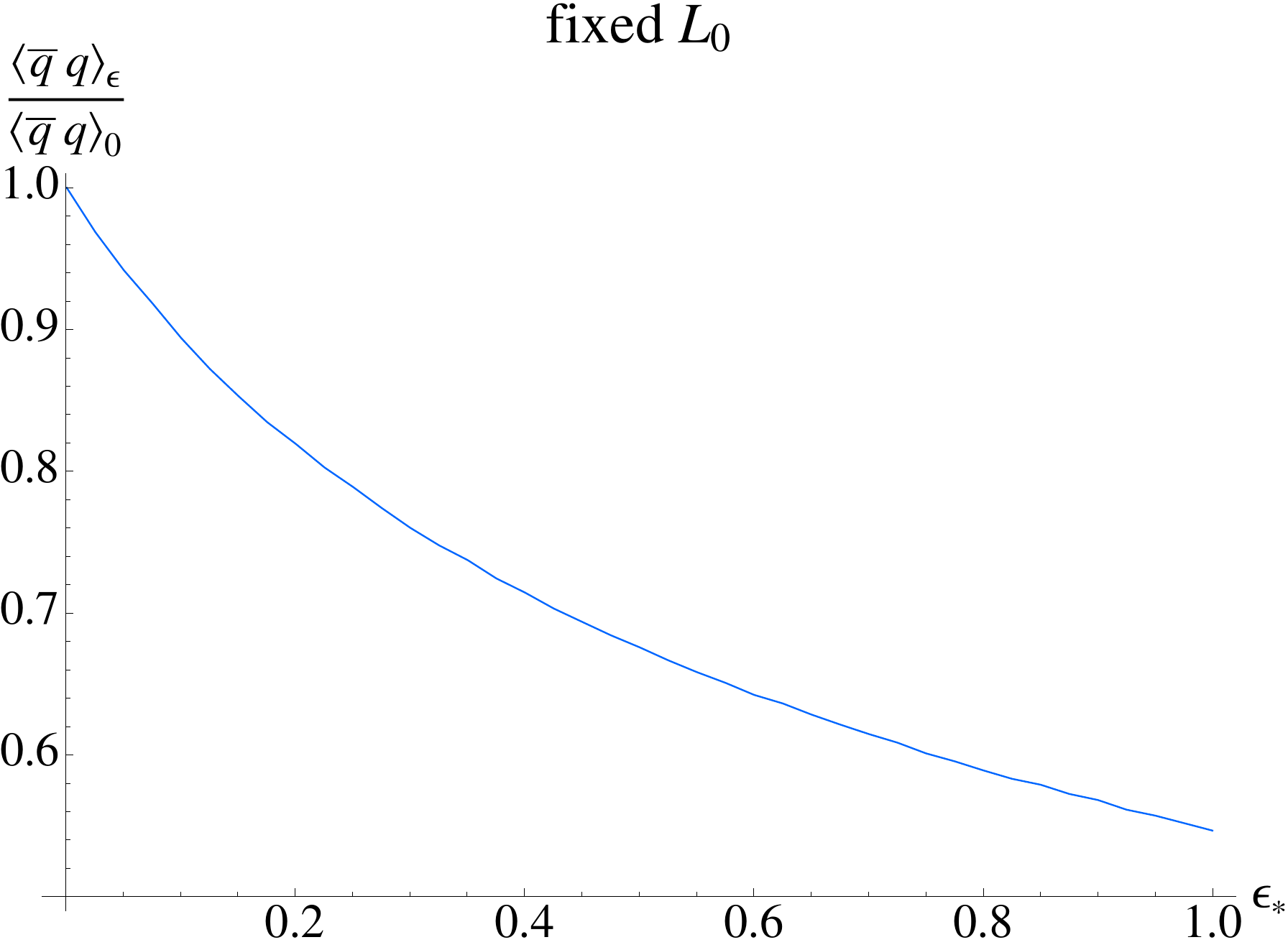} 
   \caption{\small Plots of the dependence of the scaled fundamental condensate $\langle\bar q q\rangle_{\epsilon_*}/\langle\bar q q\rangle_0$ versus the small parameter $\epsilon_*$. The left plot corresponds to varying $\epsilon_*$ at fixed $r_m$ (fixed magnetic field) and the right plot corresponds to fixed infrared separation $L_0$ (fixed constituent mass).}
   \label{fig:-.5}
\end{figure}

The first plot corresponds to a very small value of $\epsilon_*$, namely $10^{-3}$. One may expect that for 
such a small value the qualitative and quantitative behavior of the system would be the same as for the case of  
pure AdS$_5\times S^5$ space-time, studied in \cite{Filev:2007gb}. Indeed our results confirm that. The other 
plots in figure \ref{fig:-.5} correspond to $\epsilon_*=0.25, 0.5, 0.75$. One can observe that qualitatively the 
theory has the same properties. For sufficiently low bare mass (strong magnetic field) there are multiple 
phases forming a spiral structure near the origin of the $-\tilde c$ versus $\tilde m$ plane. Only the lowest 
positive ($\tilde m >0$) branch of the spiral corresponds to a stable phase \cite{Filev:2007qu}. Furthermore at 
vanishing bare mass the theory has a non-vanishing negative condensate which spontaneously breaks a 
global $U(1)$ R-charge symmetry. As one can see the only effect of the backreaction is to lower the value of 
this condensate. 

In figure \ref{fig:-.5} we have presented a plot of the symmetry breaking condensate (at vanishing bare mass) 
as a function of the parameter $\epsilon_*$. We have normalized the plots in units of the condensate at 
vanishing~$\epsilon_*$. 
Note that one can vary $\epsilon_*$ either at fixed $r_m$ (fixed magnetic field) or at fixed infrared separation  
$L_0=r_{min}$ (the radial distance above the origin at which the $S^3$ wrapped by the probe brane shrinks). 
As one can see in both cases the fundamental condensate decreases as we increase $\epsilon_*$. Note that 
in figure \ref{fig:-.5} we keep the value $\epsilon_*$ between zero and one, even if it is possible to extend 
this range. The reason behind this choice is the separation of scales, which keeps the finite cut off well below the Landau pole only in the limit $\epsilon_*\ll1$.


\section{Probing the fully backreacted background}

In this section we introduce a probe D7-brane to the perturbative background obtained in section 2. On the 
field theory side this corresponds to an introduction of additional fundamental flavors. The fundamental fields 
couple to the background $B$-field, which on the field theory side corresponds to an external magnetic field. 
Therefore, in the limit of vanishing bare mass we expect to observe the phenomenon of magnetic catalysis of 
mass generation according to \cite{Filev:2007gb}. The novel feature of our set up is the presence of smeared 
backreacted flavors which number is controlled by the parameter $\epsilon_*$. Our immediate goal is to study 
the effect of the presence of the sea of massless fundamental fields on the process of mass generation. 
Note that in this set up the external magnetic field couples to the smeared massless fundamental flavors. 
This is to be contrasted to the case studied in section 3.2, when the external magnetic field couples only to the 
flavor fields introduced by the probe brane. Our investigation indicates a significant change in the physical 
properties of the system.


\subsection{Classical embeddings}

Let us begin by considering a probe D7-brane extended along  the $t,x^i,\sigma,\theta, \psi, \phi$ coordinates 
and having a non trivial profile along the coordinate $\chi$. The effective action of the probe in the Einstein 
frame is
\begin{equation}
{\cal S}=-T_7\int d^8xe^{\Phi}\sqrt{-det(\hat g+e^{-\Phi/2}B_2)}+T_7\int P[ C_{(8)}-B_{(2)}\wedge C_{(6)}]\ ,
\end{equation}
where $C_{(8)}$ and $C_{(6)}$ are background Ramond-Ramond forms sourced by the smeared flavor branes. One can show that the effective lagrangian is
\begin{eqnarray}
{\cal L}^{(0)}&\propto& \frac{1}{8}e^{\Phi}b^2S^6F^2\cos^3\frac{\chi}{2}\left(\cos^2\frac{\chi}{2}+\frac{S^2}
{F^2}\sin^2\frac{\chi}{2}\right)^{1/2}\left(1+\frac{\chi'^2}{4b^2S^6F^2}\right)^{1/2}\left(1+\frac{e^{-\Phi}H^2h}
{b^2}\right)^{1/2} 
\nonumber \\
&+&\frac{Q_f}{32}e^{2\Phi}b^2S^8\left(1+\frac{e^{-\Phi}H^2h}{b^2}\right)\cos^4\frac{\chi}{2}\ .
\label{classLAG}
\end{eqnarray}
We find it convenient to introduce the radial variable $r$ related to $\sigma$ via \eqref{sigma-to-r}. 
In order to obtain numerical solutions for the classical embedding, we define as usual the following 
dimensionless variable
\begin{equation}
\tilde r=\frac{r}{r_m}\ ,\quad {\rm with} \quad r_m\equiv\frac{e^{-\Phi_*}Q_c H_*^2}{4} \, .\label{dmlsr}
\end{equation}
It is convenient to write the dimensionless action in terms of the following auxiliary functions
\begin{eqnarray}
\tilde{\cal L}^{(0)}=-f_1(\tilde r)\left(\cos^2\frac{\chi}{2}+f_2(\tilde r)^2\sin^2\frac{\chi}{2}\right)^{\frac{1}{2}}\sqrt{1+f_3(\tilde r)^2\chi'^2}+f_4(\tilde r)\cos^4\frac{\chi}{2}\ ,\label{dmllag}
\end{eqnarray}
where
\begin{eqnarray}
&&f_1(\tilde r)=\frac{1}{8}e^{\tilde\Phi}\tilde b^2\tilde S^6 \tilde F^2 
\left(1+\frac{e^{-\tilde\Phi}\tilde H^2}{\tilde r^4\tilde b^2}\right)^{\frac{1}{2}}
\Big|\frac{\partial{\tilde \sigma}}{\partial\tilde r}\Big| \, ,
\quad f_2(\tilde r)=\frac{\tilde S}{\tilde F} \, , 
\nonumber \\
&&f_3(\tilde r)=\left(2\tilde b \tilde S^3\tilde F\Big|\frac{\partial{\tilde \sigma}}{\partial\tilde r}\Big|\right)^{-1} \, ,
\quad f_{4}(\tilde r)=\frac{\epsilon_*}{32}e^{2\tilde\Phi}\tilde b^2\tilde S^8
\left(1+\frac{e^{-\tilde\Phi}\tilde H^2}{\tilde r^4\tilde b^2}\right)\Big|\frac{\partial{\tilde \sigma}}
{\partial\tilde r}\Big| \, ,
\end{eqnarray}
and $\tilde\Phi\ , \tilde b\ ,\tilde S\ ,\tilde F\ ,\tilde H\ ,\tilde\sigma$ are related to the functions defined in equations (\ref{sigma-to-r})-(\ref{f-s}) via
\begin{eqnarray}
&&\tilde\Phi(\tilde r)=\Phi(\tilde r r_m)-\Phi_* \, , \quad 
\tilde b(\tilde r)=b(\tilde r r_m) \, ,
\quad \tilde S(\tilde r)=S(\tilde r r_m)/r_m \, ,
\nonumber \\
&&\tilde F=F(\tilde r r_m)/r_m \, , \quad
\tilde H(\tilde r)=H(\tilde r r_m)/H_* \, ,\quad
\tilde\sigma(\tilde r)=\sigma(\tilde r r_m)r_m^4  \, . 
\label{dimlessfn}
\end{eqnarray}

One can show that a smooth solution to the equations of motion derived from (\ref{dmllag}) valid near the point  
that the $S^3$-sphere, wrapped by the probe brane, shrinks ($\chi(\tilde{r}_{min})=\pi$) is
\begin{equation}
\chi(\tilde r)=\pi-\sqrt{a(\tilde r-\tilde r_{min})}\ ,\quad a=\frac{8f_1f_2}{f_3(-2f_4+f_1f_3f_2'+f_2(f_3f_1'+f_1f_3'))}\Big|_{\tilde r=\tilde r_{min}}\label{approx}
\end{equation}
The approximate analytic solution (\ref{approx}) can be used to fix the boundary conditions for a numerical 
shooting technique at $\tilde r=\tilde r_{min}+\delta$ for small values of $\delta$.

\begin{figure}[h] 
   \centering
   \includegraphics[width=3.2in]{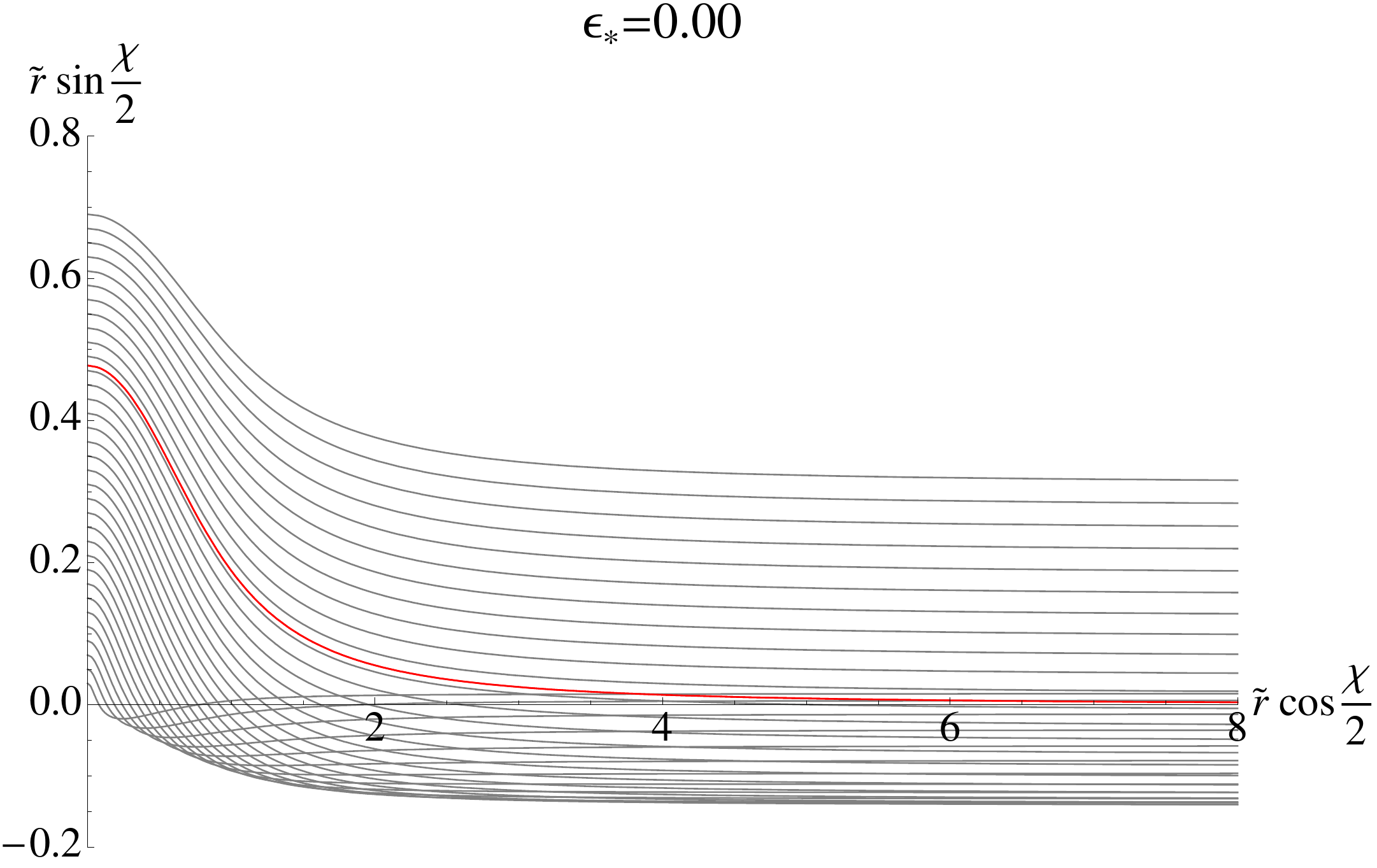} 
   \includegraphics[width=3.2in]{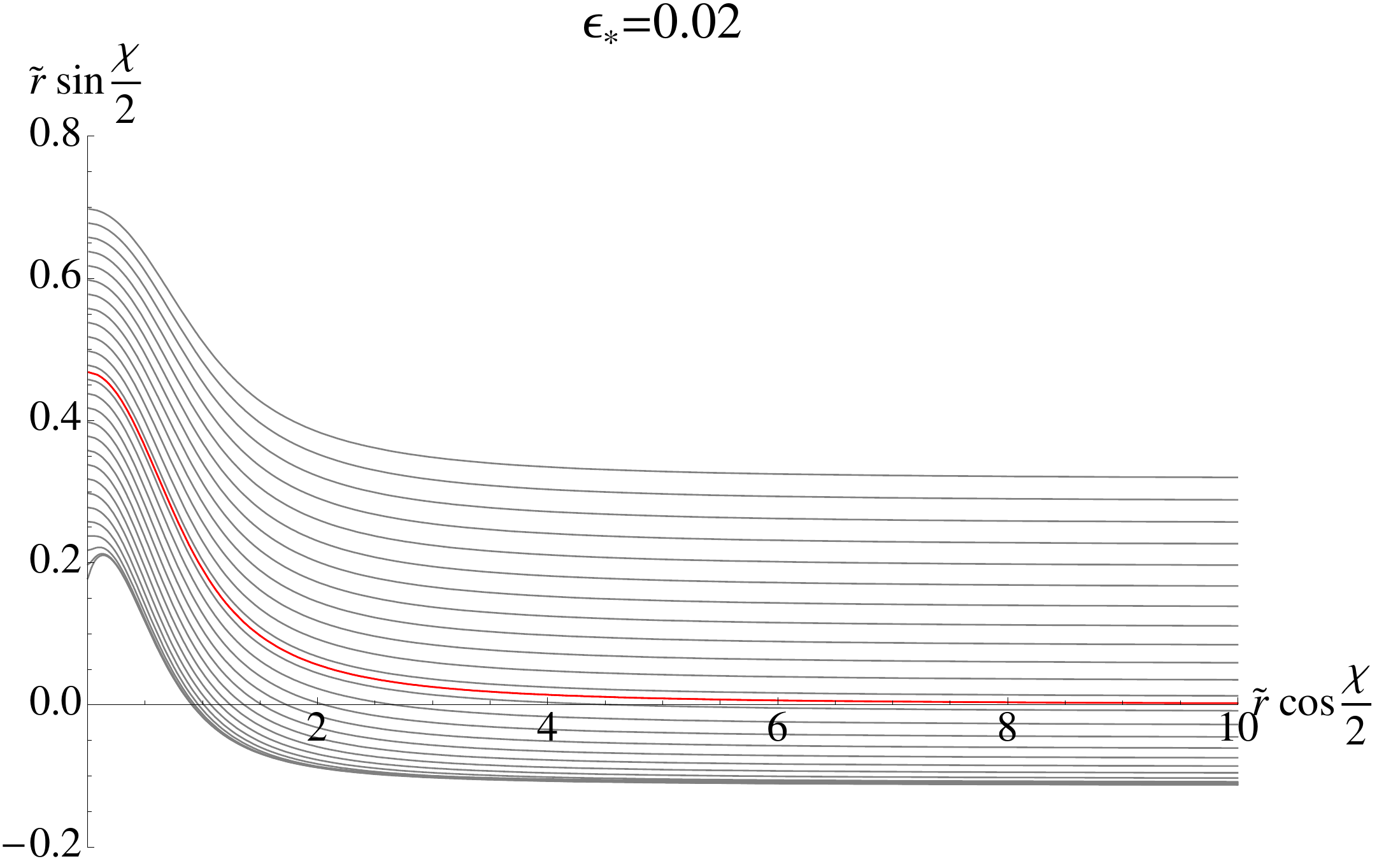} 
   \includegraphics[width=3.2in]{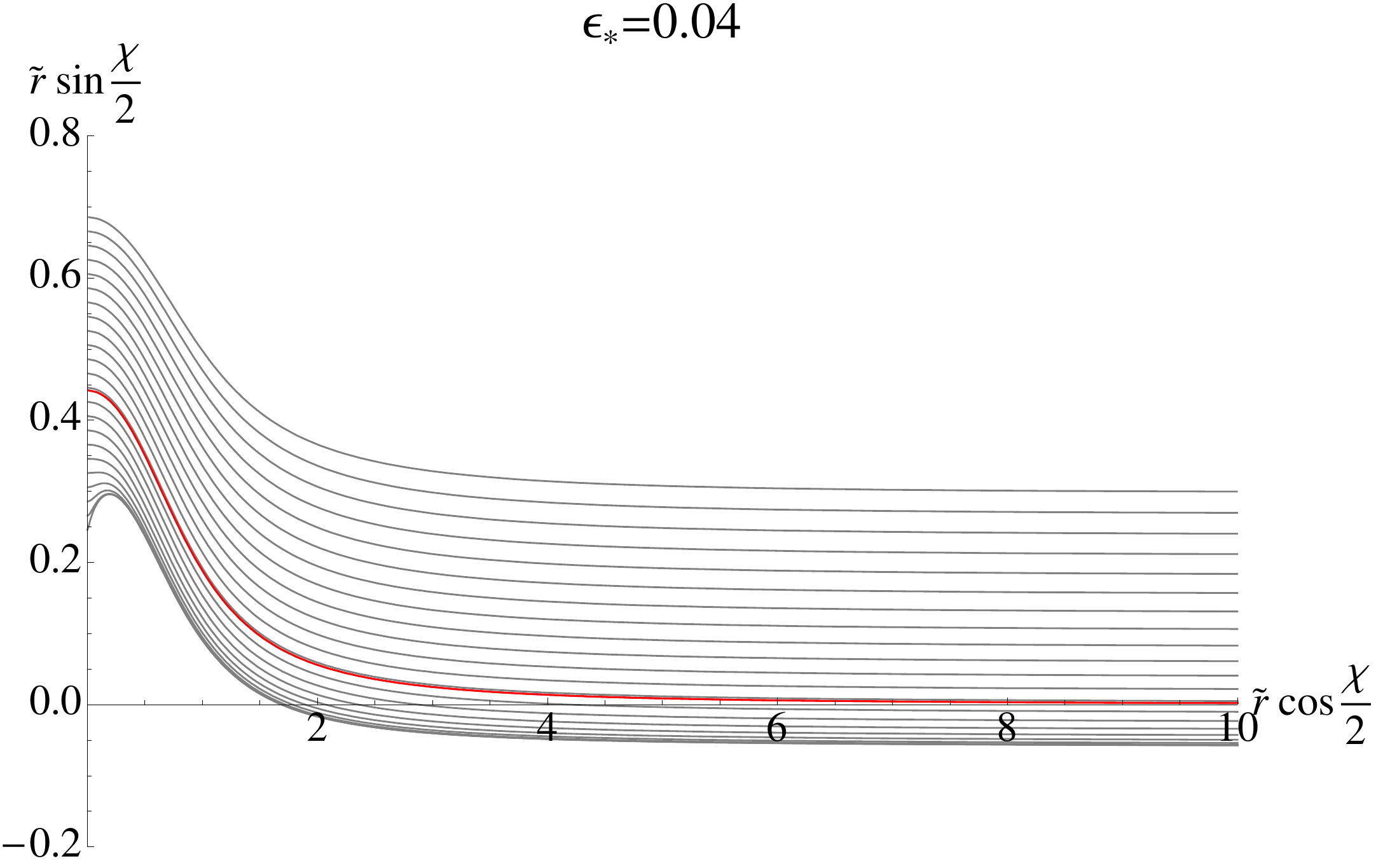} 
   \includegraphics[width=3.2in]{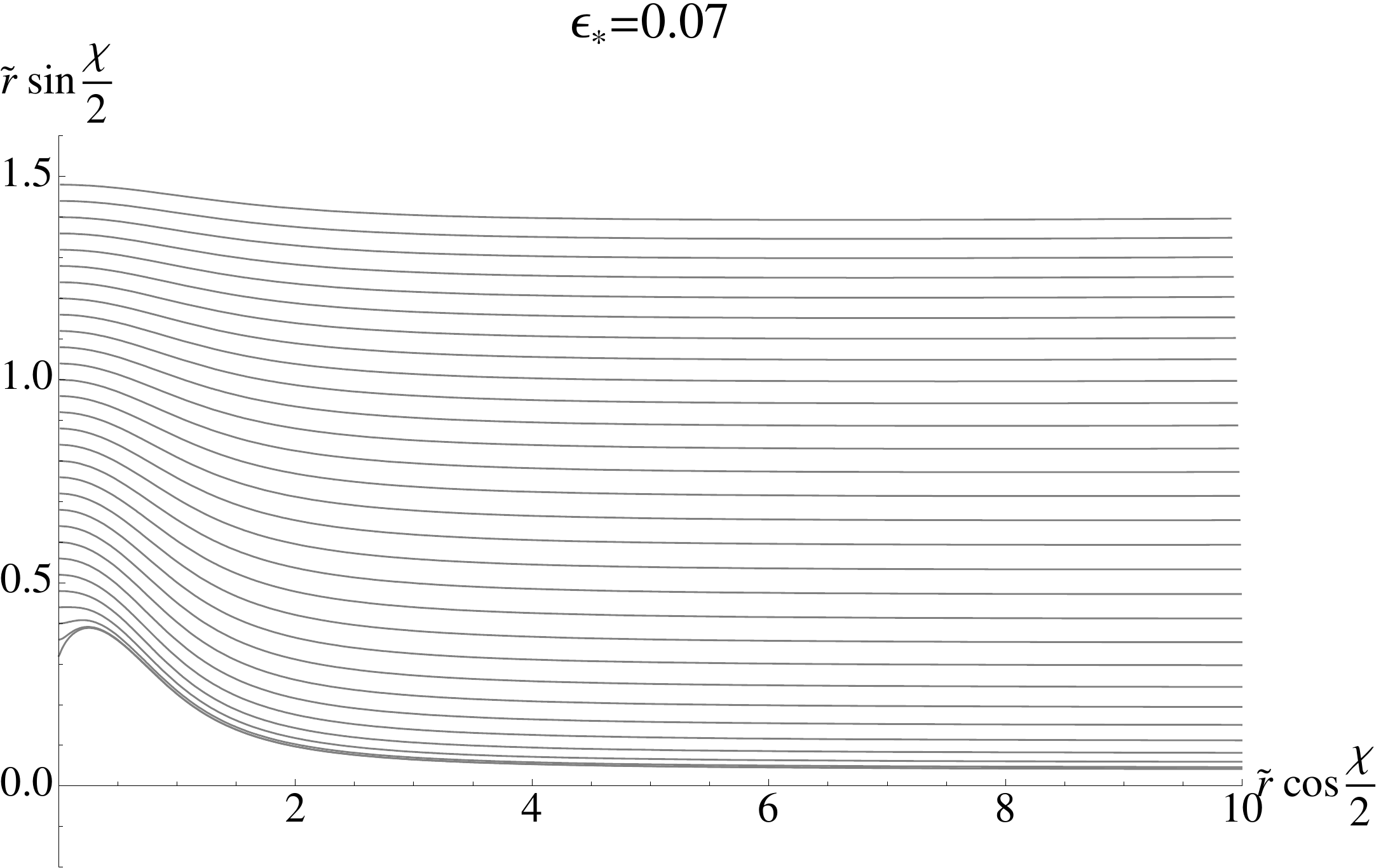} 
   \caption{\small Plots for the profiles of D7--brane embeddings having different infrared separations for 
   various values of the perturbative parameter $\epsilon_*$. The red curves correspond to embeddings with 
   vanishing separation at $r=r_*$, which we interpret as describing fundamental fields with vanishing bare 
   mass and non-vanishing constituent mass.}
   \label{fig:1}
\end{figure}

A crucial point is that due to the infrared divergency of the perturbative solution, \eqref{sigma-to-r}-\eqref{F-S}, 
this is not valid below a certain radial distance $\tilde r_{IR}(\epsilon_*)$ above the origin ($\tilde r=0$). 
The probe reflects that by having its tension vanishing at $\tilde r_{IR}(\epsilon_*)$ but in fact it is the 
Jacobian $\Big|\frac{\partial{\tilde \sigma}}{\partial\tilde r}\Big|$ which is vanishing at this value of the radial 
coordinate. As one may expect if we increase the number of backreacted flavors the radius $\tilde r_{IR}
(\epsilon)$ grows. Therefore we start loosing some of the probe brane embeddings corresponding to low bare 
masses. Fortunately this effect is somewhat softened by the effect of magnetic catalysis of mass generation 
which leads to infrared separation of the probe branes (the probe closes at some distance above the horizon). 
To visualize this behavior in figure \ref{fig:1} we have presented plots of D7-brane embeddings in $(r,\chi)$ 
polar coordinates for a range of different infrared separations ($\tilde L_0=\tilde r_{min}$) and different number 
of backreacted flavors~($\epsilon_*\propto \lambda_* \, \frac{N_f}{N_c}$).

The first plot corresponds to the absence of backreacted matter studied in \cite{Filev:2007gb}. 
As one can see there are two classes of embeddings: 
those which asymptote to negative (positive) separation at large $\tilde r$. 
One can show that \cite{Filev:2007gb} only the second class corresponds to stable embeddings. 
The two classes are separated by a critical embedding (depicted with a red curve in figure \ref{fig:1}), 
which has zero separation at large $\tilde r$ (vanishing bare mass) and finite separation in the 
infrared (dynamically generated mass).

The second and the third plots correspond to small (non-vanishing) values of the parameter $\epsilon_*$. 
As one can see due to the infrared divergencies some of the embeddings with small separation in the 
infrared are lost. However the critical embedding is still present and we can study how its separation in the 
infrared (related to the dynamically generated mass) depends on the number of backreacted flavors.

The last plot in figure \ref{fig:1} corresponds to a sufficiently large number of $\epsilon_*$, at which the radius 
of the region which is not accessible by the probe brane has increased so much that the embedding with the 
lowest separation in the infrared asymptotes to positive separation at large $\tilde r$.

At first sight it seems that the validity of the perturbative background is not sufficient to detect a qualitative 
change of the theory as a function of the parameter $\epsilon_*$. Furthermore unlike the case of the 
supersymmetric background studied in section 3, we do not have an analogous holographic renormalization 
procedure. This is why we cannot justify the validity of the AdS/CFT dictionary and in particular 
(\ref{dictionary}).

\begin{figure}[h] 
   \centering
   \includegraphics[width=6.4in]{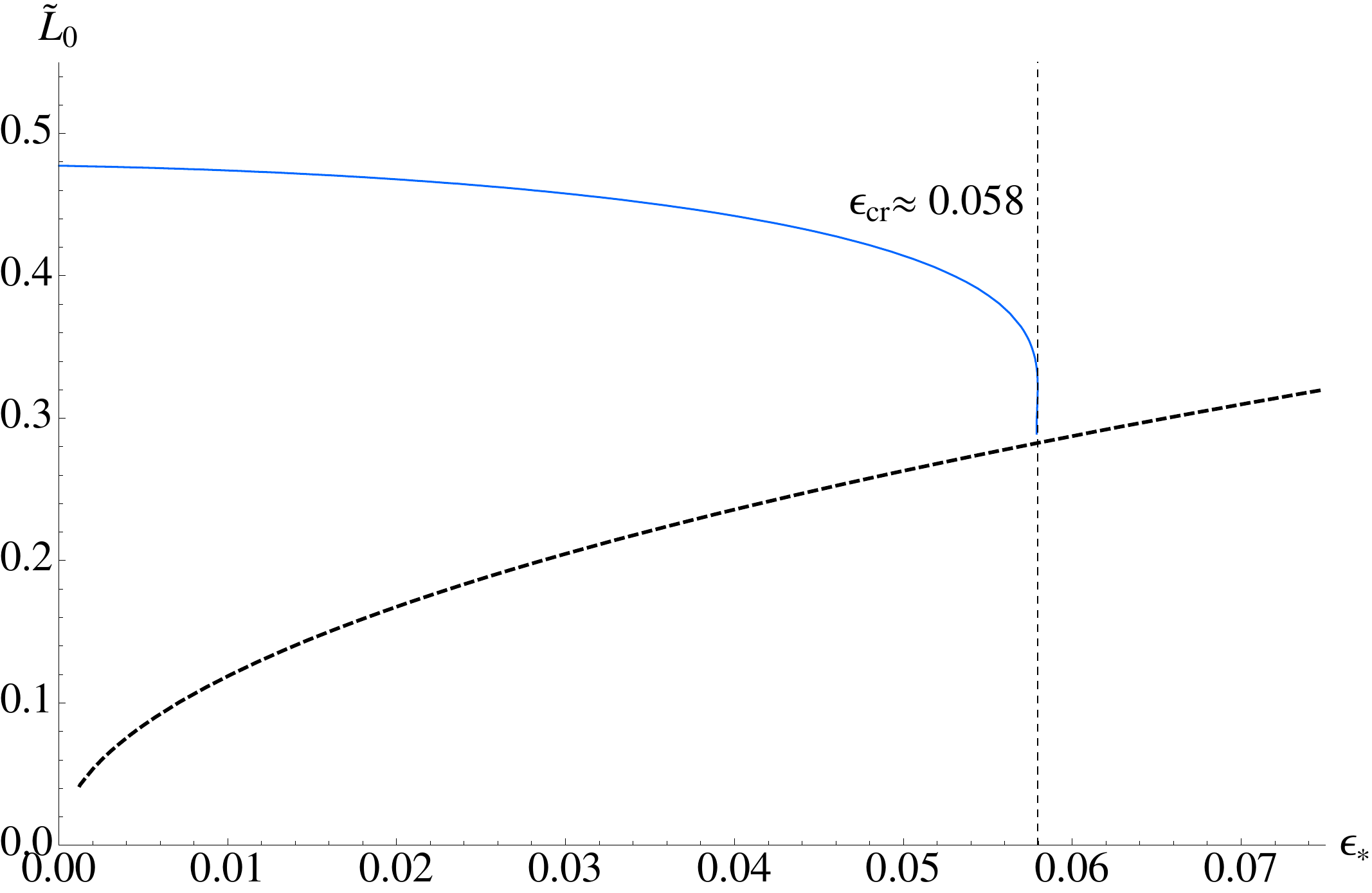} 
 \caption{\small A plot of the infrared separation $\tilde L_0$ (at zero bare mass) 
 versus $\epsilon_*\propto \lambda_* \, \frac{N_f}{N_c}$.}
   \label{fig:2}
\end{figure}

Fortunately one can use the properties of the D7-brane embeddings in the infrared, such as the infrared 
separation $\tilde L_0$ to study  the constituent mass of the fundamental fields $M_q$. Indeed a detailed 
study of the behavior of the infrared separation $\tilde L_0(\epsilon_*)$ of the critical embedding as a function 
of the parameter $\epsilon_*$ is presented in figure {\ref{fig:2}}. 

The blue curve corresponds to the function $\tilde L_0(\epsilon_*)$. The dashed growing curve corresponds to 
the radial distance $\tilde r_{IR}(\epsilon_*)$ below which the perturbative solution for the background 
geometry cannot be trusted. The study indicates that for sufficiently large values of $\epsilon_*$ the function 
$\tilde L_0(\epsilon_*)$ becomes multivalued through a diverging slope at $\tilde L_0\approx 0.326$. 
This in turn implies that the critical embedding becomes unstable at this point. 
The study of the meson spectrum supports that.

It is plausible to interpret the instability of the probe D7-brane as reflecting an instability of the background. 
Indeed if we consider a single ``fiducial" embedding, representative for the smeared massless flavor branes
\footnote{Note that unlike the critical embedding, the smeared massless flavor branes correspond to trivial ($
\chi\equiv 0$) embeddings, which are unstable in the presence of non-vanishing $B$-field}, such an 
embedding would be unstable in the probe limit due to its coupling to the background $B$-field 
\cite{Filev:2007qu}. Therefore we would expect that the background constructed by smearing many such 
embeddings would be unstable.

Then a natural question arises. Is there a critical value of the B-field at which the instability is triggered or the background is unstable at any non-vanishing magnetic field?  In the probe limit at zero bare mass the only physical scale is the scale corresponding to the non-vanishing $B$-field ($H_*$) therefore one would expect that the background is unstable at any $H_*>0$. On the other side in the backreacted case the dilaton field is running and the theory develops a Landau pole. This suggests that even at vanishing magnetic field the backreacted theory has a physical scale associated to the finite UV cut off needed to keep the relevant energy scales below the Landau pole. Therefore one would expect that there is a finite $H_{cr}>0$ at which the background becomes unstable.

Let us consider again the physical process described in figure \ref{fig:2}. Qualitatively the quantity $\tilde L_0$ can be thought of as proportional to the ratio of the dynamically generated mass of the fundamental fields introduced by the probe D7-brane $M_q$ and the energy scale corresponding to the external magnetic field $\sqrt{H_*}$, namely $\tilde L_0\propto M_q/\sqrt{H_*}$. At any fixed value of $\epsilon_*$ we have three independent energy scales: $M_q, H_*$ \& $H_{cr}$. In order to compare states at different values of $\epsilon_*$ we will keep the dynamically generated mass of the quarks fixed. The data in figure \ref{fig:2} suggests that as we increase $\epsilon_*$ the value of $\tilde L_0$ decreases. At fixed value of $M_q$ this is equivalent of increasing the external magnetic field $H_*$. Assuming that the critical value $H_{cr}$ varies slowly enough with $\epsilon_*$ we conclude that at sufficiently large values of $\epsilon_*\geq \epsilon_{cr}$ the external magnetic field $H_*$ is $H_*\geq H_{cr}$ and the gravitational background is unstable. It is plausible to interpret the instability of the probe D7-brane associated to the diverging slope of $\tilde L_0(\epsilon_*)$ in figure \ref{fig:2} as reflecting an instability of the gravitational background for $H_*\geq H_{cr}$. \footnote{However we should point out that though plausible our studies based only on a probe brane calculation alone are not conclusive.}


Note that the limited validity of our perturbative solution in the deep IR prevent us from directly studying the stability of the background. For example by studying the spectrum of minimally coupled scalar field. It would be interesting to address this problem at finite temperature, where the IR behavior of the supergravity solution is regularized by the non-extremal horizon. We leave such studies for future work. 

The instability of the probe D7-brane could be the sign 
of a ``new" physical effect, called ``superconducting 
vacuum"\footnote{We thank Maxim Chernodub for pointing out to us this possible interpretation.}. 
According to \cite{Chernodub:2010qx}, the quantum vacuum (i.e., an empty
space) may become a superconductor (in the usual electromagnetic
sense) under a strong enough magnetic field. This magnetic field
forces the electrically charged vector mesons to condense via a
tachyonic instability, which in turn implies electromagnetic superconductivity. 
Calculations in a variety of models, supporting this idea, 
can be found in \cite{Chernodub:2011mc, Braguta:2011hq, Callebaut:2011ab}.


\subsection{Meson spectrum}
In this section we analyze part of the spectrum of meson like excitations of the dual gauge theory. To this end we perform a semi-classical quantization of the probe D7--brane embeddings studied in section 4.1. In order to obtain the spectrum of the corresponding quantum fluctuations we expand systematically the effective action of the probe D7--brane to second order in~$\alpha'$. 

\subsubsection{Quadratic fluctuations.}

For a detailed study of the light meson spectrum of the theory we need to consider the quadratic 
fluctuations of a D7--brane and study the corresponding normal modes, \cite{Kruczenski:2003be}.  
The relevant pieces of the action are
\begin{equation} \label{Fluct-DBI+WZ}
\frac{S}{T_7}  =-  \int d^8x\, e^\Phi 
\sqrt{-\det \left[G_{ab}+e^{-{\Phi \over 2}} \left(B_{ab} + F_{ab} \right)\right]}+\int P[{\cal C}_{(8)} -F_{(2)}\wedge\,{\cal C}_{(6)}
+  \frac{1}{2}{F_{(2)}\wedge F_{(2)}\wedge {\cal C}_{(4)}} ] \, , 
\end{equation}
where ${\cal C}_{(4)}, {\cal C}_{(6)}$ \& ${\cal C}_{(8)}$ are defined by:
\begin{eqnarray}
{\cal C}_{(4)} \equiv C_{(4)} - C_{(2)} \wedge B_{(2)}\ ,\quad {\cal C}_{(6)} \equiv C_{(6)}-B_{(2)}\wedge \tilde C_{(4)}\ ,\quad {\cal C}_{(8)} \equiv C_{(8)}-B_{(2)}\wedge C_{(6)}
\end{eqnarray}
and $\tilde{C}_{(4)}$ is the magnetic dual of the RR potential $C_{(4)}$
\begin{equation}
\tilde{C}_{(4)} \, = \, - \frac{1}{32} \, Q_c \sin \theta \cos^4 {\chi \over 2} \,
d \theta \wedge d \varphi \wedge d \psi \wedge d \tau \, ,
\end{equation}
We will consider fluctuations of the form
\begin{equation}
\chi \, = \, \chi_0(\sigma)+2\pi\alpha' \, \delta \chi (\xi^{a})  \quad \& \quad 
\tau \, = \, 2\pi\alpha' \, \delta \tau(\xi^{a}) ,
\end{equation}
where the indices $a, b \,=\, 0,1, \ldots , 7$ run along the worldvolume of the D7-brane and 
expand \eqref{Fluct-DBI+WZ} to second order in $\alpha'$. Before presenting the relevant terms in this
expansion it is convenient to introduce $S$ $\&$ $J$, which are symmetric $\&$ antisymmetric matrices respectively, in the following way
\begin{equation}
||{E_{ab}^0}||^{-1}=S+J \, .
\end{equation}
The non-zero elements of those matrices are
\begin{eqnarray}
&& -S^{tt}=S^{11}=G_{11}^{-1} \, ,  \quad 
S^{22}=S^{33}=\frac{G_{22}}{G_{22}^2+e^{-\Phi} H^2} \, ,  \quad 
S^{\sigma \sigma}=G_{\sigma \sigma}^{-1}  \, ,  \quad 
S^{\theta \theta}=G_{\theta \theta}^{-1} \, , 
\nonumber \\
&&
S^{\varphi \varphi}= \frac{G_{\psi \psi}}{G_{\varphi \varphi}G_{\psi \psi} - G_{\varphi \psi}^2}\, ,  \quad 
S^{\varphi \psi}= - \frac{G_{\varphi \psi}}{G_{\varphi \varphi}G_{\psi \psi} - G_{\varphi \psi}^2}\, ,  \quad 
S^{\psi \psi}= \frac{G_{\varphi \varphi}}{G_{\varphi \varphi}G_{\psi \psi} - G_{\varphi \psi}^2}\, ,
\label{S} \\
&& J^{ab}=\frac{e^{-\frac{\Phi}{2}}H}{G_{22}^{2}+e^{-\Phi} H^2}(\delta_{3}^{a}\delta_{2}^{b}-\delta_{3}^{b}\delta_{2}^{a})\ ,\label{J}
\end{eqnarray}
with 
\begin{eqnarray}
&& 
G_{11} \, = \, g_{11}^{(0)} \, ,  \quad 
G_{22} \, = \, g_{22}^{(0)}  \, ,  \quad 
G_{\sigma \sigma} \, = \, g^{(0)}_{\sigma \sigma} \,+\, g^{(0)}_{\chi \chi} \, \chi_0'(\sigma)^2  \, ,
\nonumber\\
&&
G_{\theta \theta} \, = \, g^{(0)}_{\theta \theta}  \, ,  \quad
G_{\psi \psi} \, = \, g^{(0)}_{\psi \psi} \, ,  \quad
G_{\varphi \psi} \, = \, g^{(0)}_{\varphi \psi} \, ,  \quad
G_{\varphi \varphi} \, = \, g^{(0)}_{\varphi \varphi} \,  .
\end{eqnarray}
The second order terms in $\alpha'$ expansion of the action (\ref{Fluct-DBI+WZ}) are
\begin{eqnarray}
&&
-{\cal L}_{\delta \tau \delta \tau}^{(2)}=\frac{T_7}{2} e^{\Phi} \sqrt{-E_0}\,
\Bigg[g^{(0)}_{\tau \tau}- g^{(0) \,2}_{\varphi \tau} S^{\varphi \varphi} 
-2 g^{(0)}_{\varphi \tau} g^{(0)}_{\psi \tau} S^{\varphi \psi}
- g^{(0) \,2}_{\psi \tau} S^{\psi \psi}\Bigg]
S^{ab}\partial_a\delta \tau\partial_b\delta \tau\ ,
\nonumber \\
&&
-{\cal L}^{(2)}_{\delta \chi \delta \chi}=\frac{T_7}{2} e^{\Phi} \sqrt{-E_0}\,
g^{(0)}_{\chi \chi}\, \Bigg[1- g^{(0)}_{\chi \chi} S^{\sigma \sigma} \chi_0'(\sigma)^2 \Bigg] \,
S^{ab}\partial_a\delta \chi \partial_b\delta \chi \, 
+ \, \frac{1}{2}\, \tilde{f}(\sigma) \, \delta \chi^2\ , 
\nonumber\\
&&
-{\cal L}^{(2)}_{\delta\chi\delta\tau}={T_7}{\cal J}^a\delta\chi\partial_a\delta\tau\ ,\quad
{\cal L}_{AA}^{(2)}=\frac{T_7}{4} \,  \sqrt{-E_0} S^{aa'}S^{bb'}F_{ab}F_{a'b'} \, 
+ \, \frac{T_7}{2}\,P[{\cal C}_4] \, \epsilon^{mnop}\, F_{mn}\,F_{op}\ ,
\\
&&
-{\cal L}_{\delta \chi A}^{(2)}=f(\sigma) \, \delta \chi  \, F_{34} \quad \& \quad 
{\cal L}_{\delta \tau A}^{(2)}= \frac{T_7}{32}\, 
Q_c \, H_* \,\partial_{\sigma} \Bigg[\cos^4 \frac{\chi_0(\sigma)}{2}\Bigg]\, 
\delta \tau  \, F_{12}\ .\nonumber
\end{eqnarray}
where the indices $m,n,o,p$ run in the transverse directions of the D3-branes. The auxiliary functions $f$, $g$ 
\& ${\cal J}^a$, appearing in the above equations, are defined as follows
\begin{eqnarray}
f(\sigma) &\equiv& {T_7}\partial_{\sigma} \Bigg[e^{\frac{\Phi}{2}}\sqrt{-E_0} \,J^{23} \,
g_{\chi \chi}^{(0)}\, S^{\sigma \sigma}\, \chi_0'(\sigma)\Bigg] \, -
{T_7}\, e^{-\frac{\Phi}{2}} \, J^{23}\, \partial_{\chi} \sqrt{-E_0}  \,-\partial_{\chi}{\cal L}^{(0)}_{\rm WZ} ,
\nonumber \\
\tilde{f}(\sigma) &\equiv& \, \partial^2_{\chi}{\cal L}^{(0)}\, -{T_7}\,
\partial_{\sigma} \Bigg[e^{\Phi}  \,
g_{\chi \chi}^{(0)}\, S^{\sigma \sigma}\, \chi_0'(\sigma)\, \partial_{\chi} \sqrt{-E_0}\Bigg] \, ,\\
{\cal J}^a&\equiv&e^{\phi}S^{ab}\partial_{\chi}(\sqrt{-E_0})g^{(0)}_{b\tau}-e^{\Phi}\sqrt{-E_0}S^{aa'}\partial_\chi g_{a'b'}S^{b'c}g^{(0)}_{c\tau}-\partial_{\sigma}(e^{\Phi}S^{ab}\sqrt{-E_0}g^{(0)}_{\chi\chi})\chi'g^{(0)}_{b\tau}\nonumber\ ,
\end{eqnarray}
where  ${\cal L}^{(0)}, {\cal L}^{(0)}_{\rm WZ}$ are the complete classical lagrangian of the probe brane 
(\ref{classLAG}) and the $C_{6}$ contribution of the Wess-Zumino part respectively. 
Note that the function ${\cal J}^a$ is non-vanishing only for 
$a=\phi\ , \psi$.

Note that if one restricts the fluctuating modes to depend only on the time and the holographic directions the 
coupling term in the effective lagrangian vanish and we can consistently focus only on the fluctuations 
along $\chi$.


\subsubsection{Fluctuations along $\chi$}

We consider the ansatz
\begin{equation}
\delta\chi=e^{i\omega t}\eta({\sigma}) \, , \quad \delta\tau=0 \quad \& \quad A_{\mu}=0 \, .
\end{equation}
The only relevant part of the effective action is the term ${\cal L}_{\delta\chi\delta\chi}^{(2)}$. We find it 
convenient to use the radial coordinate $r(\sigma)$ related to $\sigma$ via equation (\ref{sigma-to-r}).  
Using the dimensionless notation defined in equations \eqref{dmlsr} and \eqref{dimlessfn} we obtain the 
equation of motion for $\eta(\tilde r)$
\begin{equation}
\partial_{\tilde r}\Bigg[\frac{\tilde{\cal L}_{\rm DBI}^{(0)}f_3^2}{(1+f_3^2\chi'^2)^2}\partial_{\tilde r}\eta\Bigg]+
\Bigg[\frac{\tilde{\cal L}_{\rm DBI}^{(0)}f_5^2}{1+f_3^2\chi'^2}\tilde\omega^2-
\left[\partial_{\chi}^2\tilde{\cal L}^{(0)}-\partial_{\tilde r}
\left(\frac{ f_3^2\chi'}{1+f_3^2\chi'^2}\partial_{\chi}\tilde{\cal
L}_{\rm DBI}^{(0)}\right)\right]\Bigg]{\eta}=0\ , \label{EQNfl}
\end{equation}
where
\begin{equation}
\tilde\omega^2\equiv\frac{Q_c}{4r_m^2}\omega^2  \, ,
\quad f_5(\tilde r)\equiv\frac{\tilde S(\tilde r)}{2\tilde r^2} \, .
\end{equation}
In order to solve numerically \eqref{EQNfl} we need to impose proper boundary conditions at 
$\tilde r=\tilde r_{\min}=\tilde L_0$. Using \eqref{approx} for the classical embedding one 
obtains the following asymptotic form of \eqref{EQNfl}
 \begin{equation}
\eta''(\tilde r)+\frac{3}{\tilde r-\tilde r_{min}}\eta'(\tilde r)+
\frac{3}{4}\frac{1}{(\tilde r-\tilde r_{min})^2}\eta(\tilde r)=0 \ .\label{asimpfl}
\end{equation}
The general solution of equation (\ref{asimpfl}) is
\begin{equation}
\eta({\tilde r})=\frac{C_1}{(\tilde r-\tilde r_{min})^{\frac{1}{2}}}+\frac{C_2}{(\tilde r-\tilde r_{min})^{\frac{3}{2}}} \, .
\end{equation}
Requiring renormalizability for $\eta(\tilde r)$
\begin{equation}   
\int\limits_{{\tilde r}_{min}}^{\tilde r_*}d\tilde r\sqrt{-\tilde E^{(0)}}\Big|\eta(\tilde r)\Big|^2<\infty \, ,
\end{equation}
and the fact that $\sqrt{-\tilde E^{(0)}}\sim (\tilde r-\tilde r_{min})$ on can see that $C_2$ should vanish. 
This suggests the following substitution in equation (\ref{EQNfl})
\begin{equation}
\eta(\tilde r)=\frac{\zeta(\tilde r)}{\sqrt{\tilde r-\tilde r_{min}}}\ ,
\end{equation}
where the function $\zeta(\tilde r)$ is analytic near $\tilde r_{min}$. The resulting equation of motion for 
$\zeta(\tilde r)$ is of the form
\begin{equation}
\zeta''(\tilde r)+\left(C_1-\frac{1}{\tilde r-\tilde r_{min}}\right)\zeta'(\tilde r)+\left(C_0+B^2\tilde\omega^2-\frac{C_1}{\tilde r-\tilde r_{min}}+\frac{3}{4}\frac{1}{(\tilde r-\tilde r_{min})^2}\right)\zeta(\tilde r)=0\ ,\label{EQNzeta}
\end{equation}
where $C_0\ ,C_1$ \& $B$ are functions of $\tilde r$ and $\chi(\tilde r)$ regular at $\tilde r=\tilde r_{min}$. Equation (\ref{EQNzeta}) has predetermined boundary conditions at $\tilde r_{min}$. Therefore one can expand $\zeta(\tilde r)$ near $\tilde r_{min}$ and solve equation (\ref{EQNzeta}) order by order. The obtained approximate solution for $\zeta(\tilde r)$ can be used to fix the boundary conditions for a numerical shooting technique. The resulting numerical solution depends on the parameter $\tilde\omega$. Imposing Dirichlet boundary condition at $\tilde r=\tilde r_*$ quantizes the spectrum of $\tilde\omega$. 

\begin{figure}[h] 
   \centering
   \includegraphics[width=6.4in]{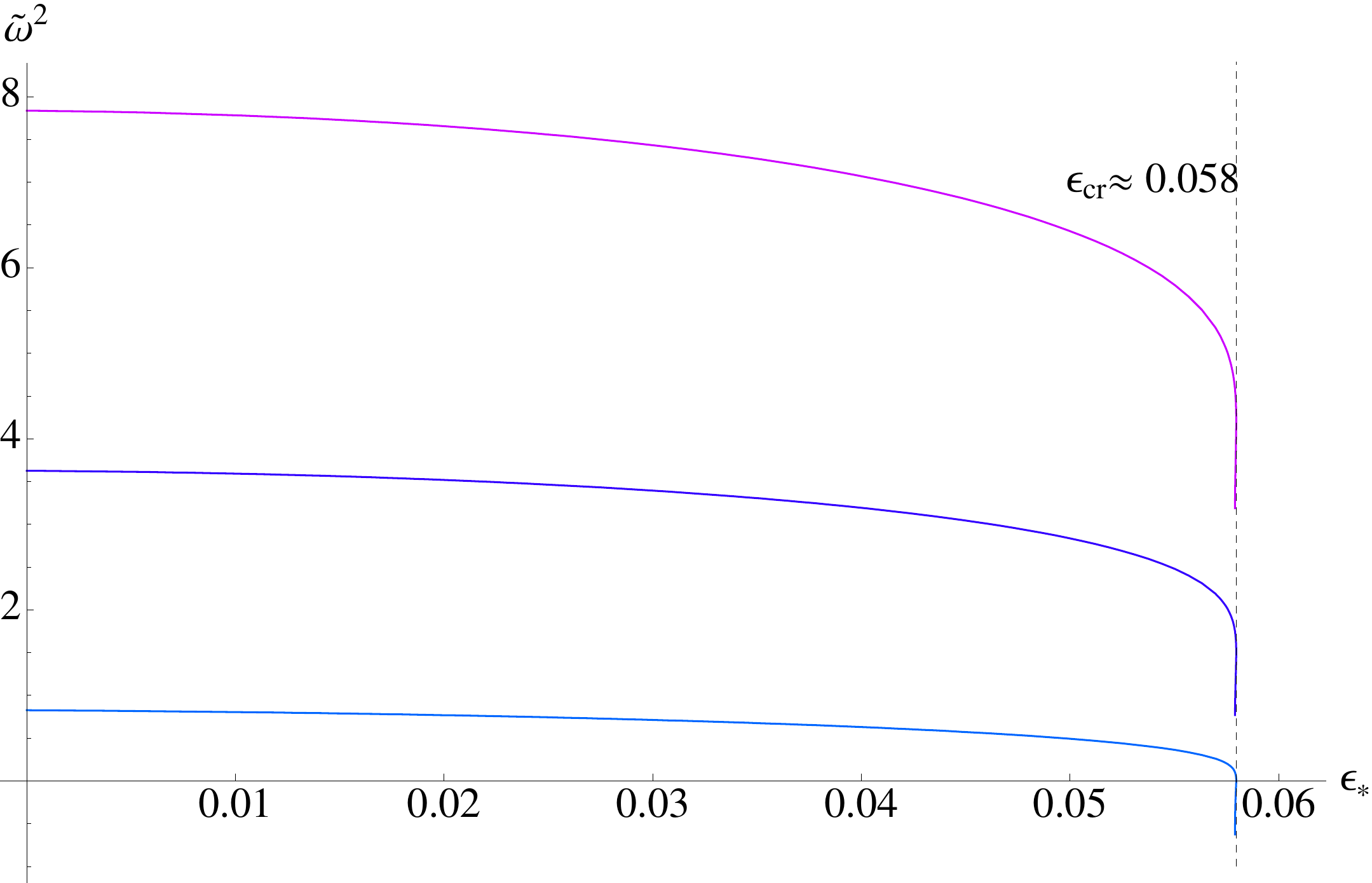} 
 \caption{\small Plots of the first three excited states of $\tilde\omega$ versus $\epsilon_*\propto N_f/N_c$. One can see that at the critical value $\epsilon_*\approx 0.058$ the ground state is tachyonic.}
   \label{fig:spect}
\end{figure}

We used the procedure outlined above to generate plots of the first three excited states of the spectrum as a function of the parameter $\epsilon_*$. The resulting plots are presented in figure \ref{fig:spect}.
One can see that the ground state becomes tachyonic exactly at the critical value of $\epsilon_*\approx 0.058$ where the slope of $\tilde L_0$ versus $\epsilon_*$ in figure \ref{fig:2} diverges. This confirms that at this point the probe branes become unstable. As we commented in subsection 4.1 we interpret this instability as reflecting an instability of the background. This suggests that the dual gauge theory is unstable for sufficiently strong magnetic fields. This is to be expected because the backreacted massless flavors are in a phase with vanishing constituent mass  which is disfavored by the external magnetic field.


\section{Conclusions}

In this paper we propose a string theory dual to a 1+3 $SU(N_c)$ ${\cal N}=4$ SYM theory coupled to 
$N_f$ massless fundamental flavors in an external magnetic field. Our motivation is to undertake the first 
steps towards an unquenched holographic description of magnetic catalysis of mass generation. By 
construction, our background corresponds to a phase of the flavored theory with vanishing constituent mass. 
This phase is disfavored by the external magnetic field, therefore one would expect that for sufficiently strong 
$B$--field the supergravity background would become unstable. Unfortunately the infrared singularity of our 
perturbative solution prevents us from directly studying the stability of the background by studying the 
spectrum of its quasi-normal modes. To circumnavigate this limitation we study the properties of an 
additional probe D7--brane.

\vspace{0.1cm}

In Section 3.1 we study the properties of a supersymmetric probe D7--brane in the limit of vanishing 
$B$--field. In this way our background reduces to the supersymmetric one obtained in 
\cite{Benini:2006hh}. We consider a holographic renormalization of the probe brane ``on-shell" action 
in the spirit of \cite{Karch:2005ms}, but in the case with backreacted flavors. Our studies reveal a remarkable 
factorization of the dependence of the ``on-shell" action on the perturbative parameter counting the number 
of backreacted flavors ($\epsilon_*$). This factorization suggests that the holographic renormalization 
performed in \cite{Karch:2005ms} can be implemented, at least formally, for the background 
\cite{Benini:2006hh}. Next, a systematic expansion in $\epsilon_*$, for $\epsilon_*\ll 1$ \& 
keeping the finite cut off of the theory sufficiently far bellow the Landau pole, provides a regime of validity of the renormalization procedure. This suggests that the usual AdS/CFT dictionary holds at the UV scale 
fixed by the finite cut off.

\vspace{0.1cm}

In Section 3.2 we move one step forward our study and introduce a non-supersymmetric probe D7--brane to 
the supersymmetric background of \cite{Benini:2006hh}. This probe brane has a fixed $U(1)$ worldvolume 
gauge field corresponding to an external magnetic field coupled to the fundamental fields introduced by the 
probe brane only. Using the AdS/CFT dictionary proposed in the previous section we study the effect of 
mass generation and its dependence on the number of backreacted fundamental flavors. Qualitatively the 
physical picture remains the same compared to the one without backreacted flavors, \cite{Filev:2007gb}. 
We contrast these results to those of a magnetic filed coupling to all fundamental degrees of freedom.

\vspace{0.1cm}

Finally in Section 4 we consider a D7--brane probe in the supergravity background obtained in Section 2. 
Now the  $B$--field, corresponding to an external magnetic field, couples to both the probe and the 
backreacted fundamental degrees of freedom. The study of the meson spectrum shows that for sufficiently 
strong magnetic field the probe brane becomes unstable. It is plausible to interpret this instability as 
reflecting an instability of the supergravity background.\footnote{However we should point out that though plausible our studies based on a probe brane calculation alone are not conclusive.} As commented above this is anticipated since the 
external magnetic field disfavors a phase with vanishing constituent mass of the fundamental fields. 
We speculate that the stable phase at vanishing bare mass would correspond to a supergravity background 
obtained by smearing ``fiducial" embeddings with a non-trivial profile along the radial coordinate having 
finite separation in the infrared corresponding to dynamically generated constituent mass. 
The construction of such a background is one of the main directions for future studies that we intend to 
pursue.

\vspace{0.1cm}

A possible way to improve the IR properties of the gravity dual constructed in Section 2 is to consider the 
non-extremal case, when the zeroth order solution is the AdS$_5\times S^5$ black hole. In this case it would 
be possible to study directly the spectrum of quasi-normal modes of the background and verify the 
anticipated instability for sufficiently strong magnetic field. One could also introduce an additional probe 
brane and perform a study analogous to the one considered in Section 4.2 of the present work. Naturally, 
one would expect that such a probe would become unstable when the energy scale of the magnetic field is 
sufficiently larger than the energy scale of the finite temperature. One can then compare the critical 
parameters obtained from studying the spectrum of quasi-normal modes of the background and from 
studying fluctuations of the probe brane. Such a study could provide an indirect check of the interpretation of 
the instability of the probe brane cosidered in Section 4.2.\footnote{We thank Aldo Cotrone for comments on this point.} We leave such studies for a future work.

\section{Acknowledgments}
V. F. and D. Z. would like to thank F. Bigazzi, M. Chernodub, A. Cotrone, J. Erdmenger, P. Kerner, I. Kirsch, S. Lin, R. Meyer, J. Mas, C. Nunez, A Paredes, A. Ramallo \& J. Shock for useful comments and suggestions. D.Z. is funded by the FCT fellowship SFRH/BPD/62888/2009. Centro de F\'{i}sica do Porto is partially funded by FCT through the projects PTDC/FIS/099293/2008 \& CERN/FP/109306/2009. The work of V. F. is funded by an INSPIRE IRCSET-Marie Curie International Mobility Fellowship. V.~F. would like to thank the organizers of the GGI Workshop "Large-N Gauge Theories" for hospitality during the final stages of this project.


\appendix

\section{Equations of motion}

The equations of motion produced from \eqref{genact} are \cite{Benini:2007kg}
\begin{eqnarray} 
R_{\mu \nu} &=& 
\frac{1}{2} \, \partial_{\mu} \Phi \partial_{\nu} \Phi + 
\frac{1}{2} \, e^{2\Phi} F^{(1)}_{\mu} F^{(1)}_{\nu} +
\frac{1}{4} \,\frac{1}{5!} \,\Bigg[ 5 F^{(5)}_{\mu \rho \sigma \alpha \beta} 
F^{(5)  \rho \sigma \alpha \beta}_{\nu} - 
\frac{1}{2} \, g_{\mu \nu} F_{(5)}^{2}\Bigg] 
\nonumber \\
&+&
\frac{1}{2}\,\frac{1}{3!} \,e^{\Phi}\Bigg[3 F^{(3)}_{\mu \rho \sigma} F^{(3)  \rho \sigma}_{\nu} - 
\frac{1}{4} \,g_{\mu \nu} F_{(3)}^{2}\Bigg] +
\frac{1}{2} \,\frac{1}{3!} \, e^{-\Phi}\Bigg[ 3 H^{(3)}_{\mu \rho \sigma} H^{(3)  \rho \sigma}_{\nu} - 
\frac{1}{4} \, g_{\mu \nu} H_{(3)}^{2}\Bigg]  
\nonumber \\
&+&
\Bigg[T_{fl} - \frac{1}{8} \, g_{\mu \nu} T_{fl}^{2} \Bigg] \, ,
\label{EOM-Einstein}
\\
{\rm d}\left[ \, \star \, {\rm d} \Phi\right] &=& e^{2\Phi} F_{(1)} \wedge \star  F_{(1)} + 
\frac{1}{2} e^{\Phi} F_{(3)} \wedge \star F_{(3)} - 
\frac{1}{2} e^{-\Phi} H_{(3)} \wedge \star H_{(3)} -
2\kappa^2 \, \frac{\delta S_{fl}}{\delta \Phi}\, , 
\label{EOM-Dilaton} 
\\
{\rm d} \left[ e^{2\Phi} \star F_{(1)}\right] & = &
- e^{\Phi} H_{(3)} \wedge \star F_{(3)}-\frac{1}{24} \, \mathcal{F}^4 \wedge \Omega_2\, , 
\label{EOM-F1}
\\
{\rm d} \left[ e^{\Phi} \star F_{(3)}\right] & = & 
- H_{(3)} \wedge F_{(5)} + \frac{1}{6} \, \mathcal{F}^3\wedge\Omega_2\, ,
\label{EOM-F3}
\\
{\rm d} \left[ \star F_{(5)}\right] & = & {\rm d} F_{(5)} = 
H_{(3)} \wedge F_{(3)} - \frac{1}{2} \, \mathcal{F}^2\wedge \Omega_2\, , 
\label{EOM-F5}
\\
{\rm d} \left[ e^{-\Phi} \star H_{(3)}\right] & = & 
e^{\Phi} F_{(1)} \wedge \star F_{(3)} - F_{(5)} \wedge F_{(3)} - 
\frac{\delta S_{fl}^{\rm\small{ (DBI)}}}{\delta \mathcal{F}}\, .
\label{EOM-H3}
\end{eqnarray}
where the last term in \eqref{EOM-H3} is an eight form and denotes the 
derivative of the smeared DBI action with respect to 
$\cal F$
\begin{equation} \label{smeared-DBI}
 \int d^8x\, e^\Phi 
\sqrt{-\det (\hat{G}+e^{-\Phi/2} \cal{F})} \quad \to \quad
 \int d^{10}x\, e^\Phi 
\sqrt{-\det (G + e^{-\Phi/2}\mathcal{F})} |\Omega_2| 
\end{equation}
The symbol $|\Omega_2|$ that appears in \eqref{smeared-DBI} denotes  the modulus of the smearing form
and has the following expression
\begin{equation}
|\Omega_2|= 2 \frac{2 Q_f}{\sqrt{h} S^{2}} \, .
\end{equation}
The Bianchi identities for all the forms of the background are
 \begin{eqnarray}
{\rm d} F_{(1)} & = & -  g_s \, \Omega_2\,  , 
\label{BIANCHI-F1}
\\
{\rm d} F_{(3)}  & = & H_{(3)} \wedge F_{(1)} - g_s \, \mathcal{F}\wedge \Omega_2 \, ,
\label{BIANCHI-F3} 
\\
{\rm d} H_{(3)}  & = & 0\, .
\label{BIANCHI-H3}
\end{eqnarray}
The Bianchi identities and the equations of motion for the worldvolume gauge fields are 
\begin{equation} 
{\rm d} \mathcal{F} = H_3 \quad \& \quad 
d \Bigg[ \frac{\delta S_{fl}}{\delta \mathcal{F}} \Bigg] = 0 \, . \label{A.12}
\end{equation}

Plugging the ansatz of \eqref{NS+RR} in the Bianchi identities
\eqref{BIANCHI-F1}, \eqref{BIANCHI-F3} and \eqref{BIANCHI-H3}  as
well as the equations of motion \eqref{EOM-F1} and \eqref{EOM-F5}, 
trivially satisfies them all. The equations of motion for $F_3$ \& $H_3$, namely 
\eqref{EOM-F3} \& \eqref{EOM-H3}, 
will give us \eqref{defJ} \& \eqref{diff-H} respectively.

The equation of motion for the dilaton, namely \eqref{EOM-Dilaton}, will give us \eqref{diff-Phi}. The 
contribution from the last term in \eqref{EOM-Dilaton} is the following
\begin{equation}
2\kappa^2 \, \frac{\delta S_{fl}}{\delta \Phi} = 
- \frac{4 e^{\Phi} Q_f}{\sqrt{h}\,S^2} \, \frac{1+
\frac{e^{-\Phi} H^2 h}{2 b^2}}{\sqrt{1+\frac{e^{-\Phi} H^2 h}{b^2}}} \, .
\end{equation}

The Einstein equations, namely \eqref{EOM-Einstein}, will give us \eqref{diff-b}, \eqref{diff-h},
\eqref{diff-S} \& \eqref{diff-F} together with the constraint \eqref{constraint}. This last one is 
coming from the $44$ (frame) component of the Einstein equations. The 
contribution from $T_{fl}$ that appears in \eqref{EOM-Einstein} is the following
\begin{equation}
T^{fl}_{\mu \nu}\,=\,\frac{2\kappa^2}{\sqrt{-g}} \, \frac{\delta S_{fl}}{\delta g^{\mu \nu}} = 
\frac{1}{2}\, g_s \, e^{\Phi} \, \sqrt{-\cal{E}} \Bigg[ {\cal S}_{\mu \nu} |\Omega_2| -\frac{2}{|\Omega_2|}\, 
\Omega^{(2)}_{\rho \mu} \Omega^{(2) \, \rho}_{\, \, \nu} \Bigg] \, ,
\end{equation}
with 
\begin{equation}
{\cal E}_{\mu \nu}\, = \eta_{\mu \nu} \, + \, e^{-\frac{\Phi}{2}}\,\frac{H\,\sqrt{h}}{b}
\left(\delta_{2}^{\mu} \delta_{3}^{\nu} - \delta_{2}^{\nu} \delta_{3}^{\mu}\right) \, , 
\quad (\mu,\nu = 0,1,\ldots ,9)
\end{equation}
and ${\cal S}$ is the symmetric part of the inverse of ${\cal E}$
\begin{equation}
{\cal S}_{\mu \nu}\, ={\rm diag} \{-1,1,\frac{1}{1+e^{-\Phi}\,\frac{H^2\, h}{b^2}},
\frac{1}{1+e^{-\Phi}\,\frac{H^2\, h}{b^2}},1,1,1,1,1,1\}\, .
\end{equation}
Finally for the worldvolume gauge fields, we have the first of the equations in \eqref{A.12} trivially
satisfied while the second splits into
\begin{equation}
d \Bigg[ \frac{\delta S_{fl}^{\rm{\small(DBI)}}}{\delta \mathcal{F}} \Bigg] +
d \Bigg[ \frac{\delta S_{fl}^{\rm{\small(WZ)}}}{\delta \mathcal{F}} \Bigg] = 0\ .\label{A.12-cont}
\end{equation}
The first term in \eqref{A.12-cont} can be easily evaluated if one differentiates \eqref{EOM-H3} and 
plugs in the ansatz \eqref{NS+RR}. 
One can show that it is of order $d\left[{\cal O}(dA)\right]$. The second term is proportional to the expression
\begin{equation}
d\left[C_{(6)}\wedge\Omega_2+{\cal O}(dA)\right] \rightarrow d\left[{\cal O}(dA)\right] \, .
\end{equation}
The bottom line is that the whole expression in \eqref{A.12} is of order 
$d\left[{\cal O}(dA)\right]$ and hence one 
can consistently set the D-brane worldvolume gauge field to zero~($A\equiv 0$).


\section{Technical details}

In this appendix we will present some technical details on the computations of Section 3.


\subsection{Exact On-Shell Action}
The ``on-shell" action $-S_{cl}/({\cal N}\alpha'^2)$ of a supersymmetric embedding with bare mass parameter $e^{\rho_q}$ in the supergravity background described by equations (\ref{SS-S})--(\ref{SS-h}) is given by
\begin{equation}
\int\limits_{\rho_q}^{\rho_*}d\rho\frac{e^{\Phi_*}(e^{2\rho}-e^{2\rho_q})[6\epsilon_*(e^{2\rho}-e^{2\rho_q})(1+
\epsilon_*(\frac{1}{6}+\rho_*-\rho))+4(1+\epsilon_*(\rho_*-\rho))(6e^{2\rho}(1+\epsilon_*(\rho_*-\rho)+\epsilon_* 
e^{2\rho_q})]}{32\times 6(1+\epsilon_*(\frac{1}{6}+\rho_*-\rho))^{\frac{1}{3}}(1+\epsilon_*(\rho_*-\rho))^2}\ .
\label{B1}
\end{equation}
The definite integral in (\ref{B1}) can be solved in a closed form using the relatively simple dependence on 
the parameter $e^{\rho_q}$. To this end one defines $m_q\equiv e^{\rho_q}$ and takes the second derivative with respect to $m_q^2=e^{2\rho_q}$
\begin{equation}
\frac{\partial^2S_{cl}[\rho_*,m_q^2]}{\partial(m_q^2)^2}=-{{\cal N}\alpha'^2}\frac{e^{\Phi_*}}{16}\left(1+\frac{\epsilon_*}{6}\right)^{\frac{2}{3}}\ .\label{B2}
\end{equation}
Furthermore equation (\ref{B1}) suggests that $S_{cl}\big|_{\rho_q=\rho_*}=0$ and one can verify that 
$\partial_{(m_q)^2}S_{cl}\Big|_{\rho_*=\rho_q}=0$. 
Therefore we can integrate \eqref{B2} and obtain the following expression for the classical action
\begin{eqnarray}
-\frac{S_{cl}}{{\cal N}\alpha'^2} = \frac{e^{\Phi_*}}{32}\left(1+\frac{\epsilon_*}{6}\right)^{\frac{2}{3}}
\left (e^{2\rho_*} -e^{2\rho_q} \right)^2 \, ,
\end{eqnarray}       
which is exactly \eqref{exactcl}.

\subsection{The Function $h$}

In order to evaluate the effective action from \eqref{effact} we need to integrate the equation of motion  for 
$h(\rho)$ in \eqref{SS-h}. Changing variables to dimensionless ones along \eqref{changevar}, 
the equation of motion for $h(\tilde\rho)$ becomes
\begin{equation}
\frac{\partial h}{\partial\tilde\rho}=-4e^{-4\tilde\rho}
\Bigg[1+\epsilon_*\left(\frac{1}{6}+\tilde\rho_*-\tilde\rho\right)\Bigg]^{-\frac{2}{3}}\ .
\end{equation}
A short integral expression can be given in terms of $x\equiv 1/\epsilon+1/6+\tilde\rho_*-\tilde\rho$
\begin{equation}
h(x)=e^{-4\tilde\rho_*}\left(1+\frac{4e^{-\frac{4}{\epsilon}-\frac{2}{3}}}{{\epsilon_*}^{\frac{2}{3}}}
\int\limits_{x_*}^{x}dx\frac{e^{4x}}{x^{\frac{2}{3}}}\right) \quad {\rm with} \quad 
x_*\equiv x(\tilde\rho_*)=\frac{1}{\epsilon_*}+\frac{1}{6} \, . \label{B2.2}
\end{equation}
The constant of integration in \eqref{B2.2} is fixed by requiring $h(\tilde\rho_*)=e^{-4\tilde\rho_*}$. 
Using that
\begin{equation}
\int dx\frac{e^{4x}}{x^{\frac{2}{3}}}=\frac{e^{i\frac{2\pi}{3}}}{2^{\frac{2}{3}}}
\Bigg[\Gamma\left(\frac{1}{3},-4x\right)-\Gamma\left(\frac{1}{3}\right)\Bigg]+\rm{const} \, ,
\label{incgamma}
\end{equation}
where $\Gamma(a)$ and $\Gamma(a,z)$ are the complete and incomplete gamma 
functions\footnote{Note that the right-hand side of equation (\ref{incgamma}) is real for $x>0$.}, we have
\begin{equation}
h(\tilde\rho)=e^{-4\tilde\rho_*}\Bigg[1+\frac{4e^{-\frac{4}{\epsilon}-\frac{2}{3}(1-i\pi)}}{{(2\epsilon_*)}^{\frac{2}{3}}}\left[\Gamma\left(\frac{1}{3},-\frac{4}{\epsilon_*}-\frac{2}{3}-4(\tilde\rho_*-\tilde\rho)\right)-\Gamma\left(\frac{1}{3},-\frac{4}{\epsilon_*}-\frac{2}{3}\right)\right]\Bigg]\ .
\end{equation}
Note that for $\tilde\rho\in (-\infty,\tilde\rho_*)$ \& $\epsilon_*>0$ $h(\tilde\rho)$ is real.

\end{document}